\documentclass[twocolumn]{aastex701}

\usepackage{amsmath,mathrsfs}

\shorttitle{Nowotka et al.}
\shortauthors{Nowotka et al.}
\graphicspath{{./}{figures/}}

\defcitealias{2003ApJS..145..329H}{HT03}
\defcitealias{2004ApJS..151..271H}{HT04}

\begin{document}

\title{Joint Analysis of HI Absorption Zeeman Measurements and the Morphology of Filamentary HI Emission}
\author[0009-0002-0282-4188]{Marta Nowotka}
\affiliation{Department of Physics, Stanford University, Stanford, CA 94305, USA}
\affiliation{Kavli Institute for Particle Astrophysics \& Cosmology, P.O. Box 2450, Stanford University, Stanford, CA 94305, USA}
\email{mnowotka@stanford.edu} 

\author[0000-0002-7633-3376]{Susan E. Clark}
\affiliation{Department of Physics, Stanford University, Stanford, CA 94305, USA}
\affiliation{Kavli Institute for Particle Astrophysics \& Cosmology, P.O. Box 2450, Stanford University, Stanford, CA 94305, USA}
\email{placeholder}

\author{Blakesley Burkhart}
\affiliation{Department of Physics and Astronomy, Rutgers University, 136 Frelinghuysen Rd, Piscataway, NJ 08854, USA} 
\affiliation{Center for Computational Astrophysics, Flatiron Institute, 162 Fifth Avenue, New York, NY 10010, USA}
\email{placeholder}  

\author{Laura Fissel}
\affiliation{Department of Physics, Engineering Physics and Astronomy, Queen's University, 64 Bader Lane, Kingston, ON, Canada}
\email{placeholder}  

\author{Tao-Chung Ching}
\affiliation{National Radio Astronomy Observatory, 1003 Lopezville Road, Socorro, NM 87801, USA}
\email{placeholder}  

\author[0000-0002-4217-5138]{Timothy Robishaw}
\affiliation{Dominion Radio Astrophysical Observatory, Herzberg Astronomy \& Astrophysics Research Centre, National Research Council Canada, P.O. Box 248, Penticton, BC V2A 6J9, Canada}
\email{placeholder}  

\author{Carl Heiles}
\affiliation{University of California, Berkeley, CA 94720, USA}
\email{placeholder}  

\begin{abstract}

We present a joint analysis of HI absorption Zeeman measurements and the morphology of filamentary HI emission to investigate the three-dimensional structure of the magnetic field in the diffuse neutral interstellar medium (ISM). Our analysis is based on the Arecibo Millennium Survey and new data from the Five-hundred-meter Aperture Spherical radio Telescope (FAST) toward radio sources 3C 75, 3C 207, and 3C 409. Toward 3C 409, we make a 4$\sigma$ Zeeman detection and infer $B_{\text{LOS}} = 9.1 \pm 1.9~\mu\textrm{G}$, in agreement with Arecibo results. We quantify the dispersion of HI filaments at the locations and velocities of Zeeman components using GALFA-HI narrow-channel emission maps. Focusing on a subsample of 42 spectrally distinct components, we find a weak but statistically significant positive correlation (Spearman $\rho$ = 0.3, $p = 0.01$) between $|B_{\text{LOS}}|$ and the circular variance of HI filament orientation angles. To examine its origin, we characterize the environments probed by HI absorption using dust emission, 3D dust maps, OH absorption, and CO emission. We find evidence that existing HI absorption Zeeman measurements trace magnetic fields that are coherent on parsec scales, probe primarily local gas ($100$–$500$ pc, often at distances consistent with the Local Bubble wall), and exhibit systematic differences in the magnitude of $B_{\text{LOS}}$. We attribute the correlation between Zeeman measurements and filamentary HI morphology to large-scale variations in magnetic field strength and/or inclination angle across different Galactic environments, which could arise due to the Local Bubble geometry or enhanced total field strength in denser regions. 

\end{abstract}

\section{Introduction}
Among the observational tracers of magnetism in the neutral interstellar medium (ISM), Zeeman splitting is the only phenomenon that directly measures the magnetic field strength \citep{1993prpl.conf..279H, 2017ARA&A..55..111H, 2019FrASS...6...66C}. In the diffuse atomic ISM, Zeeman splitting is observed via the 21 cm spin-flip transition of neutral hydrogen (HI), either in emission \citep{1989ApJ...336..808H, 1994ApJ...424..208G, 1995ApJ...442..177M}, self-absorption \citep{2022Natur.601...49C}, or absorption against a background continuum source \citep{1968PhRvL..21..775V, 2002ApJ...564..696S}. In the presence of the magnetic field, the 21-cm line splits into a linearly polarized component at the original frequency and two elliptically polarized components with frequency shifts proportional to the total magnetic field strength. Because the linearly polarized signal is expected to be weak, the Zeeman effect is detected in circularly polarized radiation \citep{2008PhDT........13R}. Moreover, since the frequency shift is much smaller than the width of the 21-cm line, the split components cannot be resolved, and the measurement reveals only the line-of-sight magnetic field strength.

Systematic surveys of Zeeman splitting remain rare due to the weakness of the polarized signal, requiring long integration times, excellent sensitivity, and a thorough understanding of polarization systematics. With 800~h of Arecibo Observatory time, \cite{2003ApJS..145..329H} conducted the largest survey to date of Zeeman measurements in 21 cm absorption against a background source, known as the Millennium Survey. The resulting sample of 66 Zeeman measurements enabled the estimation of the mean magnetic field strength in cold Galactic HI and its energy contribution relative to turbulence and thermal pressure \citep{2005ApJ...624..773H}. Additionally, the scaling of Zeeman-inferred magnetic field strength with volume density in the neutral hydrogen regime has been an important and actively debated constraint on the role of magnetic fields in molecular cloud and star formation \citep{2010ApJ...725..466C, 2015MNRAS.451.4384T, 2020ApJ...890..153J, 2023MNRAS.521.5604H, 2025MNRAS.540.2762W, 2025MNRAS.539.1024S}. The 1.4 GHz sensitivity of the now-decommissioned Arecibo Telescope was recently matched by the Five-hundred-meter Aperture Spherical radio Telescope (FAST) \citep{2018IMMag..19..112L, 2019SCPMA..6259502J, 2020RAA....20...64J, 2020Innov...100053Q}, allowing us to revisit some of the original Arecibo observations in this work.

One of the main limitations of the Zeeman measurement under typical ISM conditions stems from its sensitivity only to the component of the magnetic field vector parallel to the observer’s line of sight. As a result, for a given $B_{\text{TOT}}$, the Zeeman measurement will record a value in the range of $[0, B_{\text{TOT}}]$ depending on the inclination angle of the magnetic field with respect to the line of sight. To infer the mean value of the total magnetic field strength, previous analyses marginalized over the unknown inclination angle by assuming it is random across independent measurements \citep{2005ApJ...624..773H, 2010ApJ...725..466C}. However, to measure the total magnetic field probed by the Zeeman effect in individual spectral lines, it is necessary to obtain velocity-resolved information about the inclination angle of the magnetic field, or equivalently, about the plane-of-sky magnetic field strength, in a compatible tracer. 

In recent years, HI emission filaments have emerged as a velocity-resolved tracer of the plane-of-sky magnetic field orientation. Long, high-aspect ratio structures observed in narrow velocity channel maps of HI emission are ubiquitous over the 21 cm sky and have been shown to align with the plane-of-sky magnetic field as traced by starlight polarization \citep{2014ApJ...789...82C} and polarized dust emission \citep{2015PhRvL.115x1302C, 2015ApJ...809..153M}. Parallel alignment with the magnetic field is also observed in dust filaments at column densities below a few times $N_{\mathrm{H}} \approx 10^{21}\,\mathrm{cm}^{-2}$, with a transition to perpendicular alignment above that threshold \citep{2016MNRAS.460.1934M, 2016A&A...586A.138P, 2019ApJ...878..110F}; all sight lines studied in this work have $N_{\mathrm{H}} < 8.7 \times 10^{21}\,\mathrm{cm}^{-2}$ and a median $N_{\mathrm{H}} = 1.2 \times 10^{21}\,\mathrm{cm}^{-2}$. HI filaments preferentially trace cold neutral medium (CNM), the same HI phase that is probed by Zeeman measurements in HI absorption. The link between HI emission filaments and CNM is supported by correlations with far-infrared dust emission \citep{2019ApJ...874..171C, 2016ApJ...821..117K, 2020A&A...639A..26K}, spectral-line estimates of the CNM fraction \citep{2020ApJ...899...15M, 2023ApJ...947...74L}, similar structures observed in HI self-absorption \citep{2006ApJ...652.1339M}, and the stronger association of sodium line equivalent widths with small-scale structures in HI channel maps than with total HI column density \citep{2019ApJ...886L..13P}.

While it is well-established that HI filaments trace the plane-of-sky magnetic field orientation, their sensitivity to the magnetic field inclination has only been inferred indirectly. \cite{2019ApJ...887..136C} used the magnetically aligned filament paradigm and the assumption that gas and dust are well mixed in the diffuse ISM \citep{2017ApJ...846...38L} to create maps that model dust polarization properties based on filament orientations in HI velocity channel maps \citep[see also][]{2024ApJ...961...29H}. The coherence of HI filaments in these maps strongly correlates with dust fractional polarization from Planck, implying that filamentary HI morphology must capture the aspects of 3-dimensional magnetic field geometry that affect polarized dust emission, including the change of the plane-of-sky magnetic field orientation along the line of sight \citep{2018ApJ...857L..10C} and the inclination of the magnetic field with respect to the observer \citep{1985ApJ...290..211L, 2019ApJ...887..159H}. Indeed, the anti-correlation between the polarization angle dispersion and polarization fraction in Planck dust maps attributed to the inclination of the magnetic field \citep{2015A&A...576A.105P} is reproducible from HI filament geometry \citep{2019ApJ...887..136C}. \cite{2019ApJ...887..136C} did not explicitly model the filament inclination angle, as it is not directly measurable from HI emission maps, but instead proposed that the algorithmic and geometric detectability of filaments accounts for the effect, such that filaments may be more prominently detected when the local magnetic field is more parallel to the plane of the sky. In this paper, we explore the applicability of HI filament statistics to constraining the total magnetic field strength probed by HI Zeeman measurements.

This work investigates the correlation between HI filament statistics and the line-of-sight magnetic field strength inferred from Zeeman measurements in HI absorption. In \S\ref{sec:zeeman-datasets}, we describe the sample of Zeeman measurements used in this work and present the results of new observations obtained with FAST. In \S\ref{sec:additional-data}, we describe ancillary datasets used to examine additional factors that might influence the line-of-sight magnetic field strength. In \S\ref{sec:img-processing}, we describe the processing of GALFA-HI maps with the Rolling Hough Transform algorithm and the construction of a filament dispersion statistic. In \S\ref{sec:blos-HI-correlation}, we examine the correlation between Zeeman field strength and HI filament dispersion and evaluate its statistical significance. Sections \ref{sec:environment} and \ref{sec:structure} focus on additional information we can extract from multi-wavelength datasets about the scale, structure, and environments probed by HI Zeeman measurements. We discuss and summarize our results in \S\ref{sec:discussion} and \S\ref{sec:conclusions}. 

\section{Zeeman Effect in HI Absorption}\label{sec:zeeman-datasets}

\subsection{Observing Zeeman Splitting in HI Absorption}\label{section:zeeman-framework}
This work focuses on Zeeman measurements in HI absorption toward background radio continuum sources, including the results of the Millennium Arecibo 21 Centimeter Absorption-Line Survey \citep[][hereafter HT03 and HT04]{2003ApJS..145..329H, 2004ApJS..151..271H} and new measurements we obtained with FAST. We first outline the observational strategy for obtaining HI Zeeman measurements in absorption with a single-dish telescope. We define Stokes parameters $I$ and $V$ as $I = XX + YY$ and $V =2\,YX$, where $XX$, $YY$, $YX$ are auto- and cross-correlations of signals from two orthogonal linear receivers denoted $X$ and $Y$. Throughout this paper, antenna temperatures refer to half of Stokes $I$, and all velocities are measured in the local standard of rest frame.

Observations of the Zeeman effect in absorption proceed by simultaneously collecting Stokes $I$ and $V$ spectra toward a continuum source, followed by a position switch off the source to isolate the HI absorption profile. The antenna temperature toward a bright source with a continuum flux $T_{\mathrm{src}}$ at a line-of-sight velocity $v$ is 
\begin{equation}\label{eq:on-off}
T_{\mathrm{on}}(v)=T_{\mathrm{exp}}(v)+T_{\mathrm{src}} e^{-\tau(v)},
\end{equation}
where $T_{\mathrm{on}}(v)$ stands for the measured on-source antenna temperature, $T_{\mathrm{exp}}(v)$ is the expected 21 cm emission that would be detected in the absence of the continuum source, and $\tau(v)$ is the HI optical depth. By moving the telescope a small distance away from the source, one isolates the quantities of interest, $\tau(v)$ and $T_{\mathrm{exp}}(v)$. To account for spatial variations in 21 cm emission, spectra from multiple off-source positions are collected and used to interpolate $T_{\mathrm{exp}}(v)$.
 
The most common Zeeman fitting procedure attributes the Stokes $V$ signal to a sum of $N$ Gaussian absorption components, each with a line-of-sight field strength $B_{\textrm{LOS},n}$. The Gaussian components are obtained from a fit to the Stokes $I$ optical depth profile expressed as 
\begin{equation}
\tau(v)=\sum_{n=0}^{N-1} \tau_{0,n} e^{-4\ln2\left(v-v_{0,n}\right)^2 / \Delta v_n^2 },
\end{equation}
where $\tau_{0}$ stands for peak optical depth, $v_{0}$ denotes central velocity, $\Delta v$ is the FWHM of the line, and index $n$ iterates through the $N$ components. This decomposition is not unique, but reproducible results can be obtained by employing autonomous algorithms \citep{2015AJ....149..138L}. While not necessary for fitting Zeeman splitting in absorption, $\tau(v)$ and $T_{\mathrm{exp}}(v)$ can be used jointly to derive spin temperatures $T_s$, from which the number density $n_{\mathrm{H}}$ and non-thermal velocity dispersion $\sigma_{v, \textrm{NT}}$ of individual absorption components can be estimated. Spin temperatures are derived by fitting the equation of radiative transfer to $T_{\mathrm{exp}}(v)$ under the assumption that WNM contributes only to emission, while CNM is detected in both emission and absorption. This procedure is described in detail in \citetalias{2003ApJS..145..329H} and \cite{2018ApJS..238...14M}.

We model the Stokes $V$ spectrum in terms of left circular polarization (LCP) and right circular polarization (RCP) \citep{1990ApJS...74..437S, 2022Natur.601...49C}. We use the IEEE definition of LCP and RCP adopted by the International Astronomical Union (IAU), and define Stokes $V = \mathrm{IEEE\ RCP} - \mathrm{IEEE\ LCP}$. For a magnetic field pointing away from the observer ($B>0$), Zeeman splitting induces a shift in the frequency of a circularly polarized spectral line, $ -\Delta \nu$ for the RCP and $+ \Delta \nu$ for the LCP, where $\Delta \nu = (Z/2) × B_{\text{LOS}}$ and $Z = 2.8 \; [\mathrm{Hz} \,\mu \mathrm{G}^{-1}]$ is the Zeeman splitting factor for HI. The sign of the frequency shift reverses when the magnetic field points toward the observer ($B<0$). The opposing frequency shifts for the RCP and LCP Zeeman components manifest as a characteristic S-shaped curve in the Stokes $V$ spectrum. We model LCP and RCP optical depth profiles denoted $\tau_{\mathrm{LCP}}$, $\tau_{\mathrm{RCP}}$ by shifting the central frequency of each Gaussian optical depth component by $\pm \Delta \nu_n$ corresponding to its line-of-sight magnetic field strength $B_{\mathrm{LOS},n}$. The on-source Stokes $V$ spectrum is 
\begin{align}\label{eq:stokesVon}
V_{\mathrm{on}}(v) &= T_{\mathrm{exp, RCP}}(v) - T_{\mathrm{exp, LCP}}(v)  \nonumber\\  
&+ T_{\mathrm{src,RCP}} e^{-\tau_{\mathrm{RCP}}(v)} - T_{\mathrm{src,LCP}}e^{-\tau_{\mathrm{LCP}}(v)} \nonumber\\ 
&\approx V_{\mathrm{exp}}(v) + T_{\mathrm{src}} \left( e^{-\tau_{\mathrm{RCP}}(v)} - e^{-\tau_{\mathrm{LCP}}(v)} \right),
\end{align}
where in the last line, we assume negligible intrinsic circular polarization of the background source, i.e., $T_{\mathrm{src,RCP}} \approx T_{\mathrm{src,LCP}} \approx  T_{\mathrm{src}}$.

In analogy to Equation \ref{eq:on-off}, the first two terms of Equation \ref{eq:stokesVon} represent the circularly polarized HI emission that would be present in the absence of the continuum source and can be determined from the position switch. In contrast to $T_{\text{exp}}(v)$, spatial derivatives of $V_{\mathrm{exp}}(v)$ are expected to be undetectable due to the low signal-to-noise ratio in Stokes $V$ \citepalias{2004ApJS..151..271H}. $V_{\mathrm{exp}}(v)$ can be modeled in terms of individual components and used to measure $B_{\text{LOS}}$, but is more susceptible to instrumental effects arising from the convolution of the polarized beam structure and extended HI emission \citep{2021hai1.book..127R}. The third and fourth terms in Equation \ref{eq:stokesVon} contain the polarized optical depth spectrum, from which $B_{\text{LOS}}$ measurements in absorption are derived. Under the assumption that the velocity shift induced by Zeeman splitting is small compared to the linewidth, this term can be shown to be proportional to the product of the derivative of the Stokes $I$ spectrum and $B_{\text{LOS}}$, as adopted in the Millennium Survey analysis \citepalias{2004ApJS..151..271H}.

The Stokes $V$ spectrum contains instrumental contributions we did not express explicitly in Equation \ref{eq:stokesVon}. The on-axis leakage arises due to the uncertainty in the measurement of the Mueller matrix coefficients and is modeled as a percentage of the Stokes $I$ signal that contributes additively to the Stokes $V$ signal; this is because the leakage from Stokes $I$ to $V$ dominates the leakages from Stokes $Q$ and $U$ to $V$. The aforementioned off-axis leakage due to the structure of the polarized beam varies between different instruments and telescope pointing positions, so it must be characterized on a case-by-case basis \citep{2001PASP..113.1247H, 2008PhDT........13R, 2025AJ....169..158C}.

\subsection{Arecibo Millennium Survey}{\label{section:Millennium}}

\begin{figure*}{}
    \centering
    \includegraphics[trim={0.3cm 0 0.4cm 0cm}, scale=0.59]{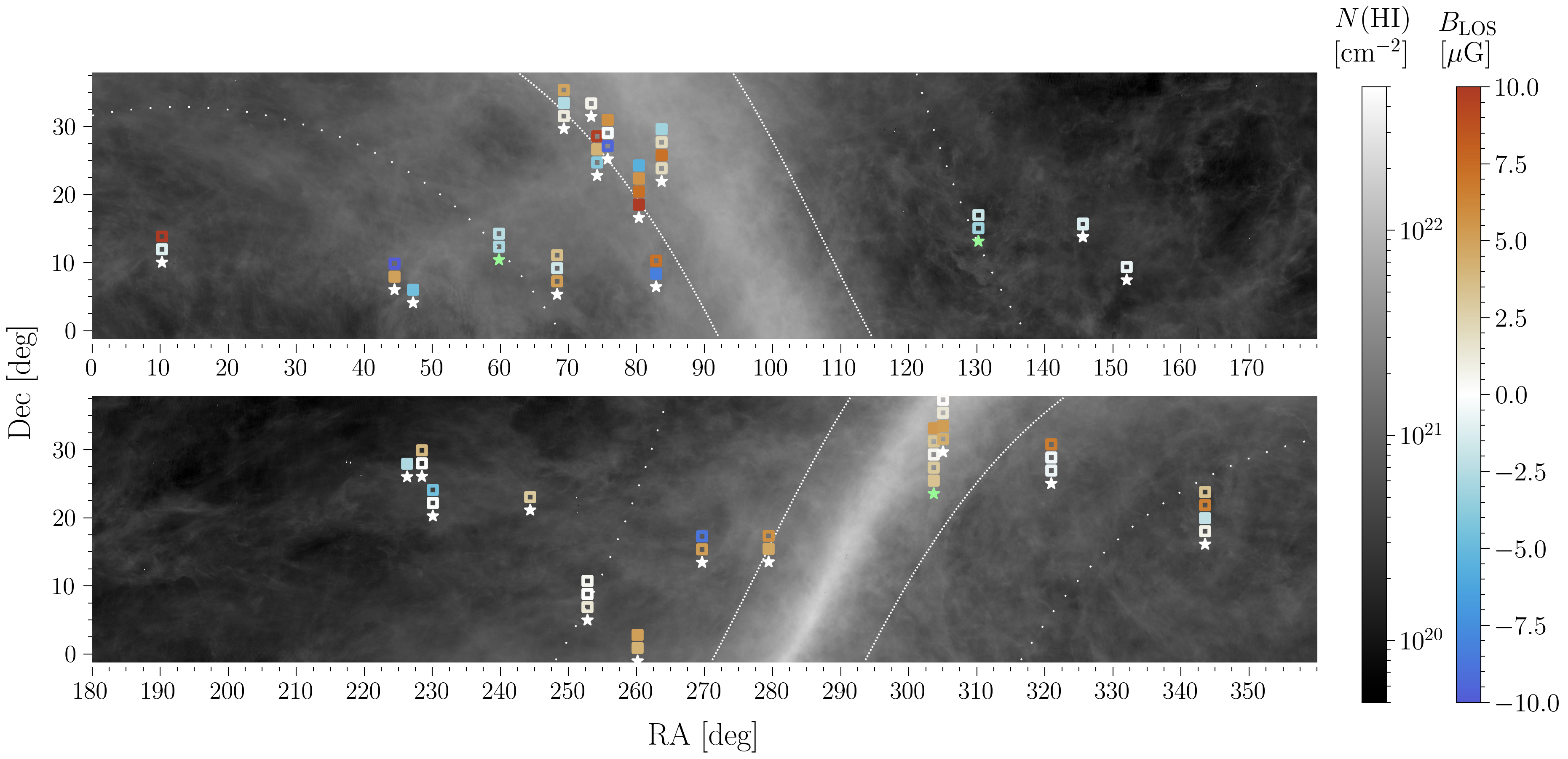}
    \caption{66 Zeeman measurements from the Millennium Survey plotted over the GALFA-HI column density map obtained for velocity range $|v|\leq 90 \,\textrm{km}\,\textrm{s}^{-1}$. Locations of the background radio sources are marked with stars, white for Arecibo measurements and green for targets where new data was collected with FAST. Zeeman measurements represented by squares are color-coded by $B_{\text{LOS}}$ inferred by HT04 and stacked in increasing radial velocity order above each radio source. Positive values of $B_{\text{LOS}}$ correspond to the magnetic field pointing away from the observer. Hollow squares indicate low signal-to-noise measurements ($\textrm{SNR} < 2$). Densely-dotted and sparsely-dotted lines mark Galactic latitudes $|b| = 10^{\circ}$ and $|b| = 30^{\circ}$, respectively. 
    }
    \label{fig:zeeman-galfa}
\end{figure*}

The primary dataset used in this work is the Millennium Survey \citepalias{2003ApJS..145..329H, 2004ApJS..151..271H}, the only systematic survey of Zeeman splitting in HI absorption over the diffuse sky. The program collected absorption and emission spectra toward 79 radio continuum sources. Forty sources with detectable CNM were selected for further observations to achieve the sensitivity necessary to study Zeeman splitting. HT04 report Zeeman measurements for 69 absorption components after including 3 archival measurements from the Hat Creek Radio Observatory 85 ft telescope toward 3C 461 (Cassiopeia A), which we omit from our analysis, as it is located outside the Arecibo footprint, where high-resolution GALFA-HI emission data are not available. 

\citetalias{2003ApJS..145..329H} analyzed the Millennium Stokes $I$ spectra to derive the physical properties of the intervening atomic hydrogen using the strategies we described in \S\ref{section:zeeman-framework}. In this paper, we make use of the products of their analysis: the line-of-sight centroid velocity $v_0$ and FWHM line width $\Delta v$ of all components, optical depth at the line center $\tau_0$, spin (excitation) temperatures $T_s$, and line-of-sight ordering of CNM components. For WNM components, the parameters include the peak unabsorbed brightness temperature $T_{0,k}$ and the fraction of WNM gas that lies in front of CNM gas $\mathscr{F}_{k}$.

In a subsequent study, \citetalias{2004ApJS..151..271H} inferred the line-of-sight magnetic field strength $B_{\text{LOS}}$ for each Gaussian component using the methods we described in \S\ref{section:zeeman-framework}. Due to instrumental effects, only the absorption spectra were used for Zeeman measurements. After excluding components with uncertainties exceeding $10~\mu \textrm{G}$ or suspicious fitting results, the survey yielded 66 Zeeman measurements (20 measurements with $\textrm{SNR} > 2$) toward 28 unique radio sources. In Figure \ref{fig:zeeman-galfa}, we plot the locations of the radio sources observed by the Millennium Survey over the Arecibo footprint. 

While other Zeeman measurements in HI absorption exist, they probe dense, star-forming regions \citep{2000ApJ...533..271S, 2012AJ....143...32M}. This work focuses primarily on the Millennium sample and the more diffuse HI regime. The Millennium Survey targeted bright radio sources, primarily extragalactic, and selected independently of foreground Galactic structures. We note that while the distribution of bright radio sources across the Arecibo footprint is unbiased, the sensitivity of individual Zeeman measurements depends on the background source flux, line optical depth, and line width \citep{1982ApJ...252..179T}. Consequently, the 66 Millennium Zeeman measurements with uncertainties below $10~\mu \textrm{G}$ represent components with higher column density than an average Millennium absorption component. Sight lines without any good Zeeman measurements ($\sigma(B_{\text{LOS}}) < 10 ~\mu \textrm{G}$) are also implicitly excluded from the Zeeman subsample.

\subsection{FAST Zeeman Data}\label{sec:fast-data}

\subsubsection{Data Collection and Processing}
We present new HI absorption Zeeman measurements obtained with FAST toward three Millennium targets 3C 409, 3C 98, and 3C 207 conducted as part of programs PT2021$\_$0068 and PT2022$\_$0163 (PI: Carl Heiles). The observations of 3C 98 took place in Sep 2022 and March 2023 and used the OnOff and TrackingWithAngle modes for a total on-source integration time of 76 min. The observations of 3C 207 took place in Nov 2021 and June 2022 and used the OnOff and TrackingWithAngle modes, totaling 112 min on-source. The observations of 3C 409 took place in Feb 2021, Sep 2022, March and May 2023 and used the OnOff and Tracking modes; the on-source time was 90 min. Each on-source integration was followed by observing one (3C 98, 3C 409) or two (3C 207) off-source positions located $5.8'$ from the initial position. All observations were performed with the full illumination of the telescope (zenith angle $<26.4^{\circ}$). We only report on the spectra from the central beam of the FAST L-band 19-beam receiver, which has low circular polarization systematics compared to secondary beams \citep{2025AJ....169..158C}. The central beam of the receiver has an average system temperature of 24 K, an aperture efficiency of 0.63, and a half-power beam width of $2.9'$ \citep{2020RAA....20...64J}. 

The output of the feed's two linear orthogonal polarization paths was cross-correlated using the ROACH backend to produce $XX$, $YY$, $XY$, and $YX$ signals. The RHSTK\footnote{RHSTK package can be accessed under \url{https://w.astro.berkeley.edu/~heiles/}} package developed by C. Heiles and T. Robishaw was used for gain and phase calibrations of the two polarization paths, bandpass calibrations of the four correlated spectra, and polarization calibrations, to produce the Stokes $I$, $Q$, $U$, $V$ spectra. We followed the calibration procedure for measuring the Mueller matrix coefficients presented in \cite{2025AJ....170..116C}. 

The uncertainty in each spectral channel is not constant and scales linearly with the system temperature defined as $T_{\mathrm{sys}}(v) =   T_A(v) + T_R$, where $T_A(v)$ is the channel-dependent antenna temperature and $T_R$ is the receiver temperature. We emphasize that this scaling also applies to Stokes $V$, which is formed by cross-correlating $X$ and $Y$ signals, so its uncertainty reflects fluctuations in $X$ and $Y$ \citep{2017isra.book.....T}. To construct an uncertainty array for each Stokes $I$ spectrum, we estimate the r.m.s. noise using 250 off-line Stokes $I$ channels and scale it by the ratio of the channel-dependent $T_{\mathrm{sys}}(v)$ to the constant off-line $T_{\mathrm{sys}}$. Similarly, for Stokes $V$, we estimate the r.m.s. noise using 250 off-line Stokes $V$ channels and scale it by the exact same ratio. The resulting uncertainty arrays are then used to combine all observations of a given on-source or off-source position by computing an inverse variance-weighted mean in each spectral channel. For 3C 207, which was observed with two different off-source positions, we use their unweighted average to represent the off-source spectrum.

\subsubsection{Gaussian Decomposition}

In \ref{section:zeeman-framework}, we described the typical Zeeman fitting procedure, which begins by decomposing the total intensity optical depth spectrum into Gaussian components. Here we adopt the Gaussian components reported by \citetalias{2003ApJS..145..329H} to enable direct comparison with the Millennium Survey and to mitigate the impact of the limited number of off-source positions on determining the spatial gradient in $T_{\mathrm{exp}}(v)$ in our FAST observations. 

To evaluate the compatibility of FAST spectra with Millennium results, we first compare the $e^{-\tau(v)}$ spectra from FAST and Millennium, finding a less than $10 \%$ difference for 3C 98 and 3C 207, and less than $15 \%$ difference for 3C 409, in each spectral channel. We attribute the observed differences to the method of $T_{\mathrm{exp}}(v)$ construction. The Millennium Survey estimated $T_{\mathrm{exp}}(v)$ by a first-order Taylor expansion fit to spectra from 16 off-source positions, while our FAST observations have at most two off-source positions. We explore whether a better agreement can be achieved with $T_{\mathrm{exp}}(v)$ based on the GALFA-HI survey maps \citep[][described further below]{2018ApJS..234....2P}. To construct $T_{\mathrm{exp}}(v)$ based on GALFA-HI, we use a linear interpolation over the absorption feature and apply a relative beam efficiency factor of $0.93$ to the interpolated emission. Using this procedure, we find a less than $10 \%$ difference from \citetalias{2003ApJS..145..329H} spectra in every channel and a mean $\tau$-weighted difference across all channels under $3 \%$ for all three targets.

We conclude that optical depth profiles derived from FAST and Arecibo spectra are consistent, justifying the use of \citetalias{2003ApJS..145..329H} Gaussian decomposition in our Stokes $V$ fitting of FAST spectra. The sparse off-source sampling is expected to have a serious effect only on the total intensity emission, since spatial gradients in polarized quantities are typically undetectable \citepalias{2004ApJS..151..271H}. While we follow \citetalias{2004ApJS..151..271H} in assuming that Stokes $V$ gradients are negligible, we recommend that future observations further confirm this assumption.

\subsubsection{Zeeman Splitting Fitting}

\begin{figure*}
    \centering
    \includegraphics[width=1\linewidth]{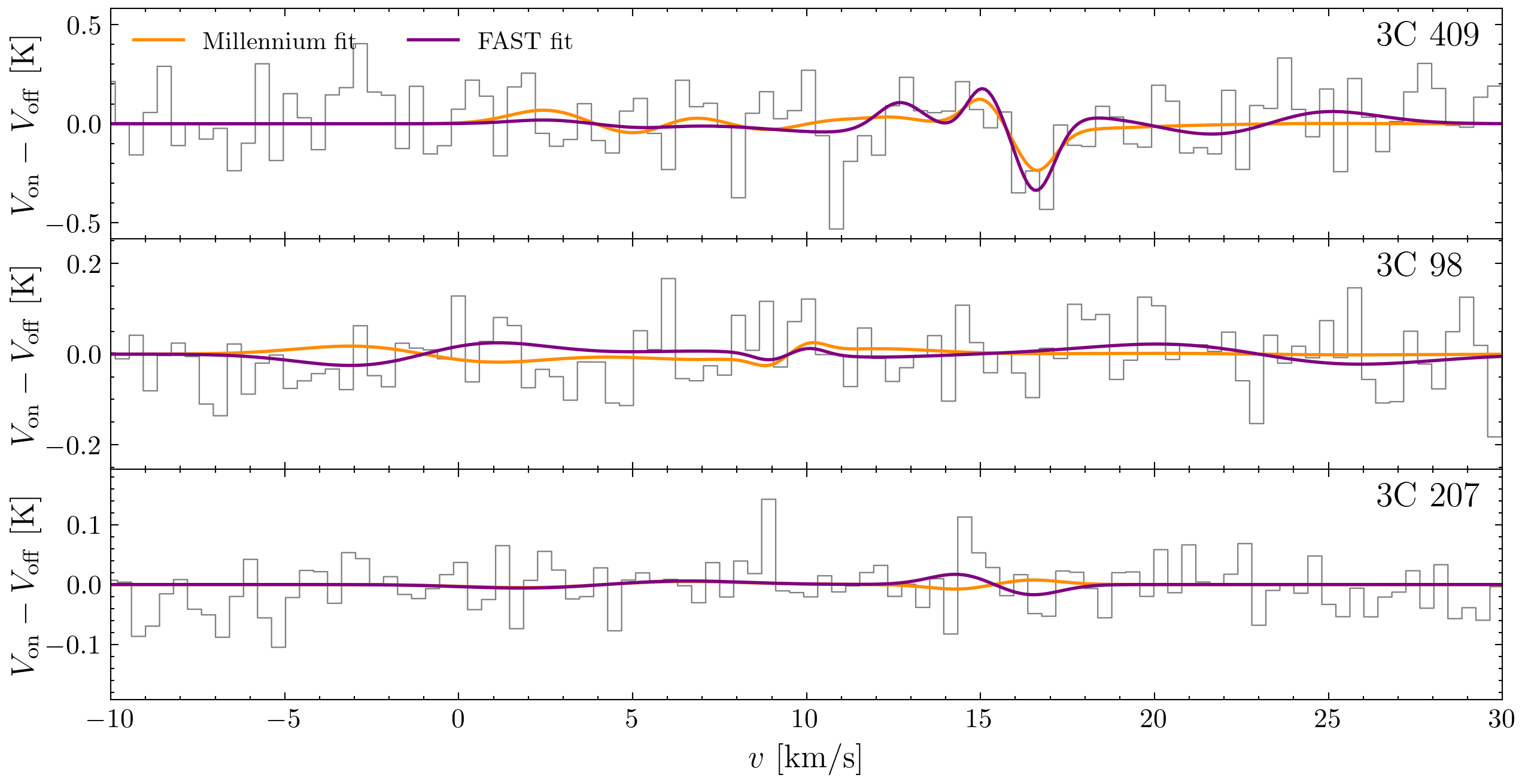}
    \caption{On-Off Stokes $V =$ IEEE RCP - IEEE LCP spectra from FAST toward sources 3C 409, 3C 98, and 3C 207 with the Zeeman splitting fit in purple. For comparison, the Zeeman splitting fit to Millennium Survey data from the Arecibo Observatory obtained by \citetalias{2004ApJS..151..271H} is plotted in orange. For visual clarity, the leakage of Stokes $I$ into Stokes $V$ has been subtracted and the spectra are binned by 4 spectral channels using the unweighted mean (each original FAST channel has a $0.1 \,\textrm{km}\,\textrm{s}^{-1}$ width and all fits to the data are performed at the original resolution).}
    \label{fig:stokesv-spectra}
\end{figure*}

To estimate $B_{\text{LOS}}$ for each absorption component, we perform a least-squares fit weighted by channel noise to the equation
\begin{align}\label{eq:zeeman-fit}
V_{\mathrm{on}}(v) - V_{\mathrm{off}}(v) & = T_{\mathrm{src}}\Bigl(e^{-\tau_{\mathrm{RCP}}(v)} - e^{-\tau_{\mathrm{LCP}}(v)}\Bigr) \\
&+ C_{\mathrm{on}}T_{\mathrm{on}}(v) - C_{\mathrm{off}}T_{\mathrm{off}}(v) \notag
\end{align}
where constants $C_{\mathrm{on}}$, $C_{\mathrm{off}}$ represent the leakage of Stokes $I$ into Stokes $V$. The left-hand side represents the observed on-source and off-source spectra and the right-hand side is the fitted curve that contains the quantity of interest $B_{\text{LOS},n}$ of each component inside the $\tau_{\mathrm{RCP}}$ and $\tau_{\mathrm{LCP}}$ terms. We do not fit for $T_{\mathrm{src}}$, which is known with high precision from Stokes $I$ observations. We have experimented with more complex fitting methods, including fitting each observation individually to account for possible leakage variations, but found nearly identical results. The leakage levels we obtain in the fits range between $0.02\% $ and $0.2\%$. Table \ref{table:fast-zeeman} includes the fitted values of $B_{\text{LOS}}$ with uncertainties. 

\begin{figure}
    \centering
    \includegraphics[width=0.97\linewidth]{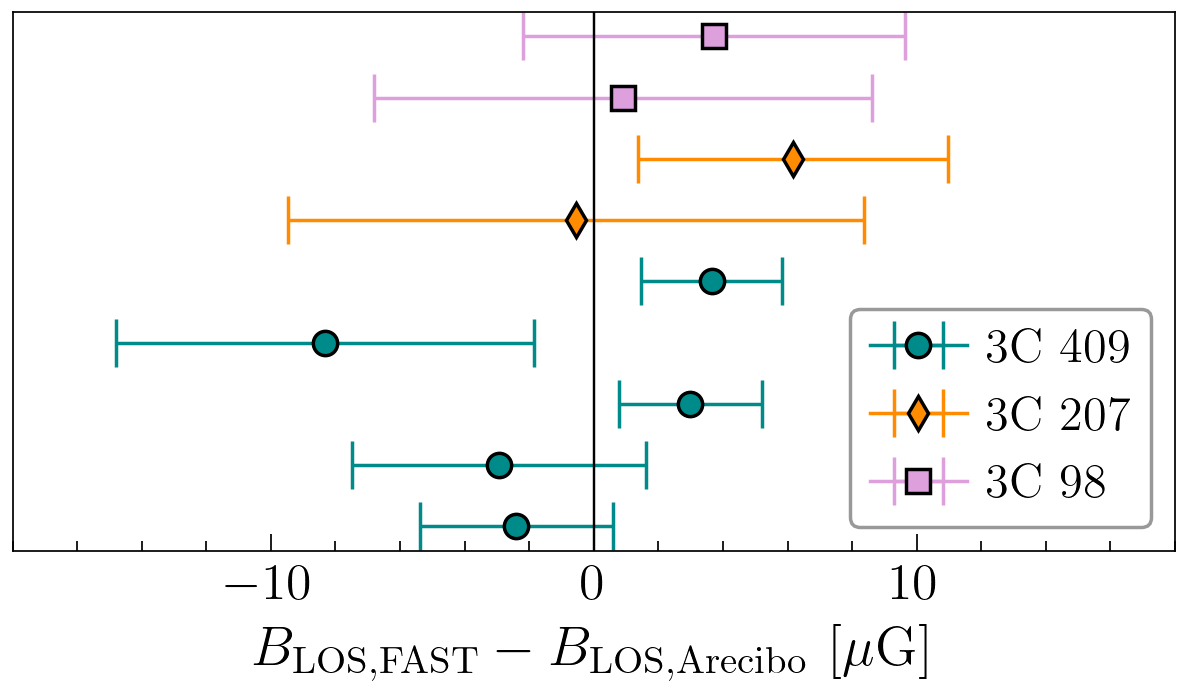}
    \caption{Difference between $B_{\text{LOS}}$ from FAST and Millennium Survey for absorption components with Millennium uncertainties $< 10~\mu\textrm{G}$. The error bars represent $1 \sigma$ uncertainties computed as $(\sigma(B_{\text{LOS,FAST}})^2 + \sigma(B_{\text{LOS,Arecibo}})^2)^{1/2}$. The Zeeman measurements from FAST and Arecibo agree within $2 \sigma$ in all absorption components. }
    \label{fig:fast-mil-comp}
\end{figure}

\begin{table}[]
\caption{Comparison of Zeeman measurements from the Millennium Survey with the Arecibo Observatory and from FAST observations presented in this paper.}
\centering

\begin{tabular}{lcccc}
\hline
Source & $v$ & $\tau$ & $B_{\mathrm{LOS,FAST}}$ & $B_{\mathrm{LOS,HT04}}$ \\
\hline
3C 98  &  -1.0 & 0.09 & $-20.8 \pm 15.9$  & $14.7 \pm 15.8$ \\
    &  9.5 & 0.30 & $-1.9 \pm 6.0$  & $-2.8 \pm 4.8$ \\
      &  9.5 & 0.51 & $\hphantom{-}1.3 \pm 4.5$  & $-2.4 \pm 3.8$ \\
      &  23.0 & 0.04 & $\hphantom{-}56.3 \pm 39.4$  & $4.2 \pm 41.4$ \\
      
3C 207 & 4.2 & 0.25 & $ -3.8 \pm 8.1$  & $-3.2 \pm 3.7$ \\
      & 15.4 & 0.30 & $ \hphantom{-}4.3 \pm 4.3$  & $-1.9 \pm 2.2$ \\
      
3C 409 &  3.9 & 0.45 & $\hphantom{-}0.9 \pm 2.8$  & $\hphantom{-}3.3 \pm 1.2$ \\
      &  7.7 & 0.33 & $\hphantom{-}0.1 \pm 4.2$  & $\hphantom{-}3.0 \pm 1.7$ \\
      & 13.7 & 0.86 & $\hphantom{-}3.6 \pm 2.0$  & $\hphantom{-}0.6 \pm 1.0$ \\
      & 14.0 & 0.41 & $-5.1 \pm 6.0$  & $\hphantom{-}3.2 \pm 2.5$ \\
      & 15.7 & 0.75 & $\hphantom{-}9.1 \pm 1.9$  & $\hphantom{-}5.4 \pm 1.0$ \\
      & 23.3 & 0.04 & $-42.7 \pm 35.0$  & $\hphantom{-}1.1 \pm 13.7$ \\

\hline
\end{tabular}

\label{table:fast-zeeman}
\end{table}

Figure \ref{fig:stokesv-spectra} shows the On-Off Stokes $V$ spectra from FAST toward 3C 409, 3C 98, and 3C 207, calibrated to Stokes $V$ = IEEE RCP – IEEE LCP following the IAU convention. The plotted best fit curves represent the first two terms on the right-hand side of Equation \ref{eq:zeeman-fit}, either with $B_{\text{LOS}}$ obtained in this work from the new FAST observations (``FAST fit"), or with $B_{\text{LOS}}$ reported by the Millennium Survey (``Millennium fit"). In Figure \ref{fig:fast-mil-comp}, we compare our Zeeman measurements from FAST with the Millennium Survey results. All FAST measurements agree with Millennium measurements within $2 \sigma$, and we reproduce the $5 \sigma$ Arecibo detection ($4 \sigma$ in FAST) in the $v = 15.7 \, \textrm{km}\,\textrm{s}^{-1}$ component of 3C 409. Our FAST measurements had shorter integration times than Arecibo observations and have modestly larger uncertainties. In subsequent analysis, when available, we use an inverse-variance weighted mean of Zeeman measurements from both instruments. However, the statistical significance of the results presented in the rest of the analysis remains unchanged when using only the original Millennium sample.

\section{Additional Datasets}\label{sec:additional-data}
Here we list the datasets we employ to characterize the HI emission morphology and Galactic environment associated with HI Zeeman measurements, including any additional quantities we derive from each dataset.

\subsection{HI Emission Maps}
The Galactic Arecibo L-Band Feed Array HI Survey \citep[GALFA-HI,][]{2018ApJS..234....2P} mapped the 21 cm emission from neutral hydrogen over the entire Arecibo telescope footprint covering $32 \%$ of the sky with the angular and spectral resolution of $4'$ and $0.184 \: \textrm{km\,}\textrm{s}^{-1}$, respectively. In this work, we use GALFA-HI to study the morphology of HI emission in narrow-channel maps centered on the velocities of Zeeman absorption components. A cumulative HI intensity map $I(v)$ centered on velocity $v$ with width $dv$ is constructed by adding individual GALFA-HI channels, 
\begin{equation} \label{eq:channel}
    I(v)=\int_{v-d v / 2}^{v+d v / 2} T_B\left( v^{\prime}\right) d v^{\prime},
\end{equation}
where $T_B(v)$ is the brightness temperature of the GALFA-HI map at velocity $v$.

\subsection{Properties of HI Components}
We derive additional properties of the CNM and WNM Gaussian components detected in absorption-emission spectra as part of the Millennium Survey, which we introduced in \S\ref{sec:zeeman-datasets}. 

The column density of a single $n$th CNM component detected in absorption is given by 
\begin{equation}\label{eq:cnm-column}
    N\mathrm{(HI)}_{\text {CNM},n}=C_0 T_s \int \tau_n(v) d v
\end{equation}
where $C_0=1.823 \times 10^{18} \mathrm{~cm}^{-2} \mathrm{~K}^{-1}~\left(\mathrm{km} \mathrm{~s}^{-1}\right)^{-1}$, $T_{\mathrm{s}}$
is the spin temperature, and $\tau(v)$ is the Gaussian optical depth profile of the $n$th component. 

The column density of a single $k$th WNM component is the integral of its unabsorbed emission profile,
\begin{equation}\label{eq:wnm-column}
    N\mathrm{(HI)}_{\text {WNM}, k} = C_0 \int T_{0, k} e^{-4 \ln 2\left(v-v_{0, k}\right)^2/\Delta v_k^2} d v.
\end{equation}

To obtain the total HI column density along the line of sight, including both the CNM and WNM gas, we integrate Equations \ref{eq:cnm-column} and \ref{eq:wnm-column} over all velocities and compute the sum
\begin{equation}\label{eq:tot-column}
\begin{aligned}
N\mathrm{(HI)}_{\text {total }} & =  N\mathrm{(HI)}_{\text {CNM}}+N\mathrm{(HI)}_{\text {WNM }} =  \\ &= \sum_{n=0}^{N-1} N\mathrm{(HI)}_{\mathrm{CNM}, n} 
 + \sum_{k=0}^{K-1} N\mathrm{(HI)}_{\mathrm{WNM}, k} 
\end{aligned}
\end{equation}
where $n$ iterates through components detected both in absorption and emission, and $k$ iterates through emission-only components. This procedure does not assume that the emitting gas is optically thin, nor that emission from every velocity channel is dominated by gas of a single temperature.


The width of an absorption line contains contributions from thermal broadening and turbulent broadening, such that
\begin{equation} \label{eq:turb_vel}
{\Delta v}^2 = 8 \, \textrm{ln}\,2 \left( \frac{k_B \, T_k}{m_\mathrm{H}} + \sigma^2_{v, \textrm{NT}} \right),
\end{equation}
where $\Delta v$ is the FWHM of the spectral line, $T_k$ stands for kinetic temperature,  and $\sigma_{v, \textrm{NT}}$ is the non-thermal velocity dispersion measured along the line of sight. In CNM, collisions are the dominant excitation mechanism of the 21 cm transition \citep{1958PIRE...46..240F, 2001A&A...371..698L}, so we adopt $T_k = T_s$. We calculate the turbulent velocity dispersion for each Zeeman component using Equation \ref{eq:turb_vel}. We note that MHD turbulence is generally anisotropic, and $\sigma^2_{v, \textrm{NT}}$ represents only the component of turbulent velocity projected along the line of sight. 

\begin{figure*}{}
    \centering
    \includegraphics[trim={0.5cm 1.1cm 0cm 1.2cm}, clip, scale=0.67]{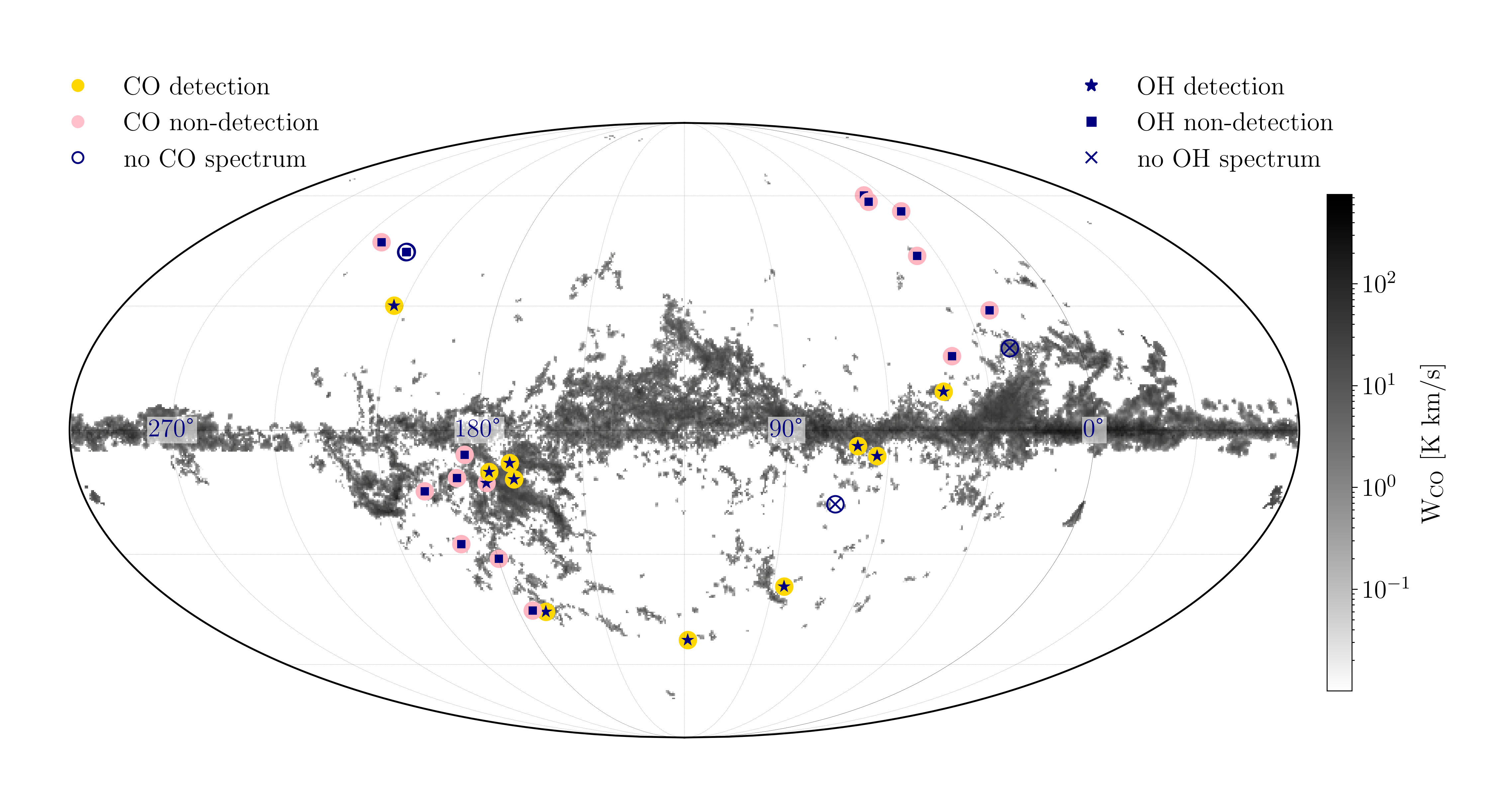}
    \caption{28 Millennium lines of sight with Zeeman measurements plotted over a velocity-integrated map of CO emission intensity from \cite{2022ApJS..262....5D} centered at $(l,b) = (120^\circ,0^\circ)$. Circles indicate CO emission line detections (yellow) and non-detections (pink), while dark blue stars and squares mark detections and non-detections, respectively, of OH absorption lines at any velocity along the line of sight, as reported by \cite{2018ApJS..235....1L}. Crosses represent lines of sight without OH emission spectra, while dark blue empty circles represent the lack of CO absorption spectra in \cite{2018ApJS..235....1L}.}
    \label{fig:zeeman-oh-co}
\end{figure*}

\subsection{CO and OH Spectra}

The Millennium Survey collected OH absorption measurements at  1665.402  and  1667.359 MHz toward 72 out of 79 Millennium sight lines, with a spectral resolution of $0.068 \mathrm{\:km \,
s}^{-1}$ and r.m.s. sensitivity of 28 mK. This dataset was first made public by \cite{2018ApJS..235....1L}, who selected 44 of the Millennium sources with OH spectra for follow-up observations in CO emission. The Purple Mountain Observatory Delingha (PMODLH) telescope of the Chinese Academy of Sciences was used to obtain J=1-0 $\text {transitions of } { }^{12}\mathrm{CO},{ }^{13} \mathrm{CO} \text {, and } \mathrm{C}^{18} \mathrm{O}$. The Caltech Submillimeter Observatory (CSO) and the IRAM 30 m telescope were used to obtain $^{12}$CO (J=2-1) transitions. Depending on the instrument and the observed transition, the sensitivity of the CO observations ranged between 0.25 to 0.6 K in a $0.16 \mathrm{\:km \,
s}^{-1}$ channel, and the angular resolution ranged between $11''$ to $50''$. In this paper, we use the derived properties of OH and CO spectral lines reported by \cite{2018ApJS..235....1L}, which include the central velocity, FWHM width, optical depth or peak brightness, and column density of each component. 

Out of 28 Millennium sight lines with Zeeman measurements, 24 and 26 sight lines have CO and OH spectra in \cite{2018ApJS..235....1L}, respectively. We show the sight-line coverage of this dataset in Figure \ref{fig:zeeman-oh-co}. For the 4 sight lines not selected for CO observations by \cite{2018ApJS..235....1L}, we use the CO (J=1-0) survey of the entire northern sky conducted with the 1.2 m Millimeter-Wave Telescope at the Center for Astrophysics $|$ Harvard \& Smithsonian \citep{2022ApJS..262....5D}. The CfA survey has a sensitivity of 0.18 K and a spectral and angular resolution of $0.65 \mathrm{\:km \, s}^{-1}$ and $0.25^{\circ}$, respectively. 

\subsection{Dust Emission Maps}\label{sec:planck}
We use Planck maps of dust reddening across the sky \citep{2016A&A...596A.109P}, constructed by separating Galactic dust emission from the cosmic microwave background and cosmic infrared background signals using a Generalized Needlet Internal Linear Combination algorithm \citep[GNILC;][]{2011MNRAS.418..467R}. The resolution of the Planck GNILC maps ranges from $5'$ in regions of bright dust emission (65\% of the sky) up to $20.8'$ where the signal-to-noise ratio of Galactic dust emission is low. We extracted $E(B-V) = \tau_{353} \times (1.49 \times 10^{-4} \, \mathrm{mag})$ at the location of Millennium radio sources using the \texttt{dustmaps} package \citep{2018JOSS....3..695G}, which queries the products of fitting a thermal dust model to GNILC maps at $5'$ resolution. We further convert $E(B-V)$ to optical $V$-band extinction using 
\begin{equation}
    A_V = R_V \times E(B-V)
\end{equation}
where $R_V = 3.1$ \citep{1989ApJ...345..245C}.

\subsection{3D Dust Maps}
To examine distances to structures probed by Zeeman measurements, we use the 3D dust map constructed by \cite{2024A&A...685A..82E} based on dust extinction and stellar distances inferred from Gaia BP/RP spectra of 54 million nearby stars by \cite{2023MNRAS.524.1855Z}. The 3D dust map has an angular resolution of $14'$ and spans distances between 69 pc and 1250 pc in logarithmic bins with a resolution ranging from 0.4 pc to 7 pc. We query 12 posterior samples of the \cite{2024A&A...685A..82E} map at the coordinates of sight lines with Zeeman measurements using the \texttt{dustmaps} package \citep{2018JOSS....3..695G}. 

    \begin{figure*}{}
        \centering
        \includegraphics[trim={0cm 6.5cm 0cm 1.5cm}, scale=0.21]{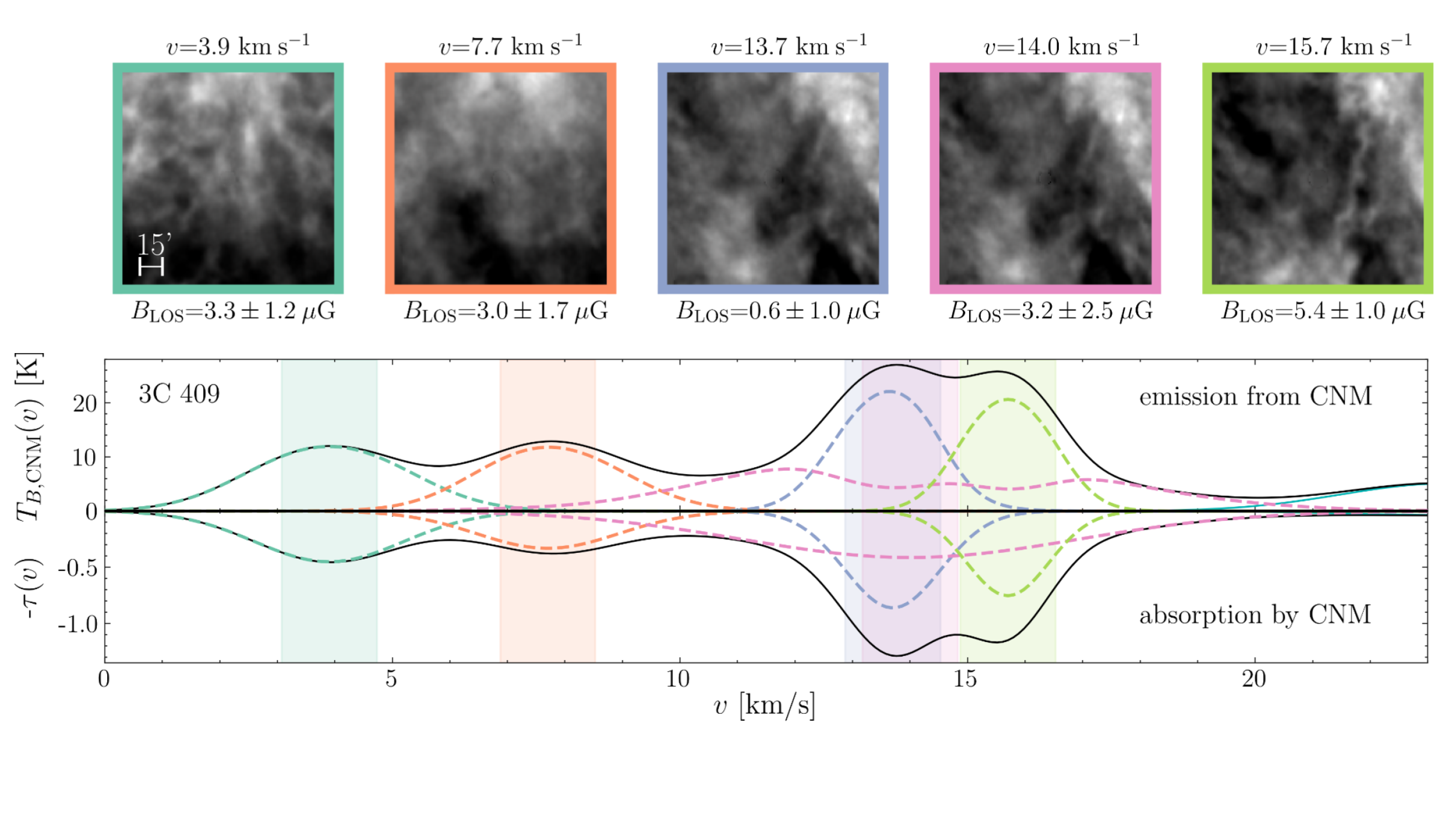}
        \caption{\textit {Top}: GALFA-HI narrow-channel emission maps centered on the radio source 3C 409 and on the radial velocities of absorption components with existing Zeeman measurements. Each map has a dimension of $2.5^{\circ} \times 2.5^{\circ}$ and a channel width of $1.7 \: \textrm{km}\,\textrm{s}^{-1}$. \textit{Bottom}: Gaussian components detected in absorption toward 3C 409 according to the \citetalias{2003ApJS..145..329H} fit. The upper plot shows the expected HI emission $T_{\mathrm{B, CNM}}$ due to each color-coded CNM component. The shaded regions represent the velocity ranges of the GALFA-HI emission maps shown in the upper panel. Emission from WNM is not shown (see \S\ref{section:crowding}).}
        \label{fig:5zeemans}
    \end{figure*}

\section{HI Filament Dispersion at the Zeeman Absorber velocities}\label{sec:img-processing}

\subsection{Preparation of GALFA-HI Maps}

We extract 66 GALFA-HI maps with a channel width of $1.7 \, \textrm{km\,}\textrm{s}^{-1}$ centered at the positions of the Millennium Zeeman sources and at the velocities corresponding to absorption components with Zeeman measurements, as illustrated in Figure \ref{fig:5zeemans}. We then implement a series of image pre-processing steps. First, we use an interpolation with a Gaussian kernel to mask the central absorption feature caused by the background radio continuum source. The spatial extent of absorption features varies between sources; the mask radius is chosen for every sight line by visual inspection and ranges from $3'$ to $10'$. We then apply a simple Fourier filtering procedure to remove regular scanning artifacts common in the GALFA-HI survey \citep{2018ApJS..234....2P}. Sections of images containing irregular scanning artifacts that are not easily removed by Fourier filtering are manually set to zero at the bitmasking stage of the Rolling Hough Transform algorithm described below. Toward 3C 98, the impact of the artifacts is severe, and 3C 353 is very close to the edge of the GALFA-HI footprint, so we exclude these two lines of sight from RHT map construction.

\subsection{Application of the Rolling Hough Transform}\label{section:RHT}
We process the GALFA-HI cutouts using the Rolling Hough Transform \citep{2014ApJ...789...82C}, an algorithm designed to detect coherent linear structures. First, an image smoothed with a top-hat filter of diameter $\theta_{\mathrm{FWHM}}$ is subtracted from the original image to highlight small-scale structures. The resulting array is thresholded such that negative pixels are set to zero. Any regions containing irregular scanning artifacts are also manually set to zero. The Hough transform is applied in a circular window of diameter $D_W$ at the location $(x,y)$ of every nonzero pixel, and quantifies the intensity of linear features passing through that pixel, $R(x, y, \theta)$, as a function of the orientation angle $\theta$. Next, $R(x, y, \theta)$ is offset by $Z_{\text{RHT}}$, the fraction of pixels within the circular window that produce non-zero RHT output, and further set to zero if the outcome of the subtraction is negative. We run the RHT algorithm using the \texttt{convRHT} implementation of \cite{2023ApJ...945...72A}, which utilizes a series of convolutions to speed up the procedure. 

The output of the RHT algorithm $R(x, y, \theta)$ can be normalized and thought of as a probability distribution of the orientation of linear structures passing through the pixel $(x,y)$. We store the normalization factor, which we refer to as RHT intensity $I_{\mathrm{RHT}}$, where 
\begin{equation}
    \label{eq:rht-intensity}
    I_{\mathrm{RHT}}(x, y, v) = \int R(x, y, v, \theta) d\theta,
\end{equation} and use it as a measure of the total intensity of linear structures integrated over all orientations. 

The RHT extracts filaments at a fixed angular scale, which is modulated by parameters $D_W$ and $\theta_{\mathrm{FWHM}}$. Increasing the window length $D_W$ results in the selection of longer filaments, while the smoothing parameter $\theta_{\mathrm{FWHM}}$ determines the thickness of the extracted features. Motivated by the analysis of \cite{2024ApJ...961...29H}, we initially run the RHT with $D_W=105'$, $\theta_{\mathrm{FWHM}}=5'$, and $Z_{\text{RHT}}=0.75$ and explore the impact of this choice in the later sections.

\subsection{Filament Orientation Statistics}
We follow the mathematical formalism of \cite{2019ApJ...887..136C} to define the plane-of-sky orientation of HI emission filaments $\theta_{\mathrm{HI}}(x, y, v)$. We use the normalized RHT output to construct $Q_{\mathrm{HI}}$ and $U_{\mathrm{HI}}$ maps for every pixel coordinate $(x,y,v)$,
\begin{equation}
\label{eq:synt-q}
 Q_{\mathrm{H} \mathrm{I}}(x,y,v)=I(x,y,v) \sum_\theta R(x, y, v, \theta) \cos (2 \theta) d \theta\, ,   
\end{equation}
\begin{equation}
\label{eq:synt-u}
U_{\mathrm{HI}}(x, y, v)=I(x, y, v) \sum_\theta R(x, y, v, \theta) \sin (2 \theta) d \theta \, .
\end{equation}
From $Q_{\mathrm{HI}}(x,y,v)$ and $U_{\mathrm{HI}}(x,y,v)$, we compute HI angle $\theta_{\mathrm{HI}}(x,y,v)$,
\begin{equation}
    \theta_{\mathrm{HI}}(x, y, v)=\frac{1}{2} \arctan \frac{U_{\mathrm{HI}}(x, y, v)}{Q_{\mathrm{HI}}(x, y, v)},
\end{equation}
which is defined only for pixels with nonzero RHT intensity. Unlike $Q_{\mathrm{HI}}$ and $U_{\mathrm{HI}}$, $\theta_{\mathrm{HI}}$ is independent of HI emission intensity.

To quantify the disorder of HI emission filaments detected by the RHT in a selected region of the sky, we use the circular mean 
\begin{equation}
    \bar{\theta}_{\mathrm{HI}}(v) = \frac{1}{2} \arctan{\frac{C(v)}{S(v)}}
\end{equation}
and circular variance
\begin{equation}
    \label{eq:circ-std-rht}
    \mathrm{Var}( \theta_{\mathrm{HI}}(v)) = 1 - (C(v)^2 + S(v)^2)^{1/2},
\end{equation}
where $C(v)$ and $S(v)$ are defined as 
\begin{equation}
    C(v) = \sum_{x,y} \cos({2\theta_{\mathrm{HI}}}(x,y,v))
\end{equation}
\begin{equation}
    S(v) = \sum_{x,y} \sin({2\theta_{\mathrm{HI}}}(x,y,v)),
\end{equation}
and the sums run over coordinates $(x,y)$ of all pixels within the selected region of the sky. Circular variance is a unitless quantity defined on the range [0,1].

\begin{figure}[ht]
        \centering
        \includegraphics[trim={0cm 0cm 0cm 0cm}, scale=0.68]{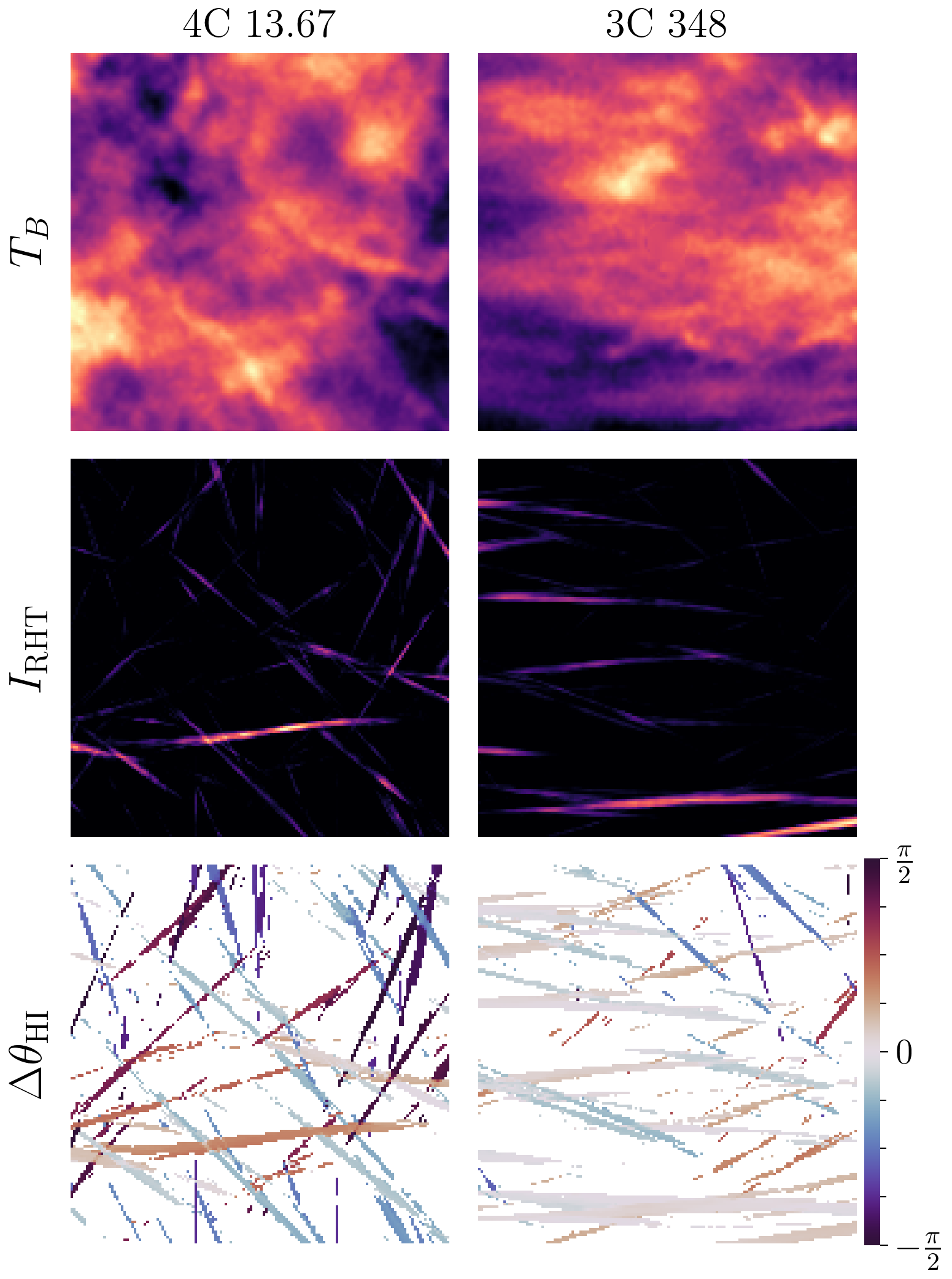}
        \caption{Top row: Examples of two $2.5^{\circ} \times 2.5^{\circ}$ GALFA-HI emission maps with  $1.7 \, \textrm{km\,}\textrm{s}^{-1}$ channel width representative of high ($\mathrm{Var}(\theta_{\mathrm{HI}}) = 0.82$, left) and low ($\mathrm{Var}(\theta_{\mathrm{HI}}) = 0.15$, right) variance of $\theta_{\mathrm{HI}}$ angles constructed using the Rolling Hough Transform. Middle row: Corresponding RHT intensity maps $R(x,y)$ quantifying the degree of linearity in every pixel of the map. Bottom row: Deviation of the RHT angle $\theta_{\mathrm{HI}}$ from the mean orientation angle in the image $\Delta \theta_{\mathrm{HI}} = \theta_{\mathrm{HI}} - \langle{\theta}_{\mathrm{HI}}\rangle$. }
        \label{fig:RHT-examples}
    \end{figure}

In Figure \ref{fig:RHT-examples}, we show examples of two emission maps with high and low variance of HI filament orientation, $\mathrm{Var}(\theta_{\mathrm{HI}}) = 0.82$ and $\mathrm{Var}(\theta_{\mathrm{HI}}) = 0.17$, respectively. Filamentary structures highlighted by RHT intensity maps are detected in both images. As expected, the emission map characterized by low $\mathrm{Var}(\theta_{\mathrm{HI}})$ shows a high 
alignment of RHT angles relative to the mean orientation.

\section{Results}\label{sec:blos-HI-correlation}
In this section, we investigate the connection between Zeeman measurements in HI absorption and filamentary HI emission quantified using the Rolling Hough Transform. 

\subsection{Spectral Complexity}\label{section:crowding}
The association between HI absorption components and narrow-channel HI emission maps is non-trivial for several reasons. First, HI emission in the degree-scale GALFA-HI maps might not all be 
associated with gas seen in the pencil-beam absorption spectra. Second, about half of Millennium sources are located at $|b| < 20^{\circ}$, where multiple CNM components along the line of sight give rise to absorption lines that are not well-separated in velocity space. If these absorption components originate from spatially distinct regions, the observed narrow-channel HI emission will contain blended contributions from their distinct morphologies.

To mitigate the first issue, we report $\mathrm{Var}(\theta_{\mathrm{HI}})$ over the smallest angular scales that provide sufficient filament statistics. We address the second difficulty by selecting a subsample of absorption components that dominate the emission in the immediate vicinity of the absorption sight line, as inferred from the \citetalias{2003ApJS..145..329H} emission-absorption line pairs. The expected emission from the $n$th CNM component is
\begin{equation}
T_{\mathrm{B}, \mathrm{CNM}, n}(v)= T_{s, n}\left(1-e^{-\tau_n(v)}\right) e^{-\sum_{m=0}^{M-1} \tau_m(v)},
\end{equation}
where subscript $m$ iterates through $M$ clouds laying in front of the $n$th cloud, and the component order was determined from least-squares fit to emission spectra by \citetalias{2003ApJS..145..329H}. Our selection criterion excludes absorption components that contribute less than $70 \%$ of the total CNM emission within the $1.7 \, \textrm{km\,}\textrm{s}^{-1}$ velocity range around their centroid velocities $v_0$. For example, toward 3C 409 (Figure \ref{fig:5zeemans}), all but the $v = 14 \, \: \textrm{km}\,\textrm{s}^{-1}$ component (marked in pink) pass this criterion. Our definition is agnostic to the intensity of WNM emission because the RHT highlights narrow filamentary structures, which are preferentially associated with CNM \citep{2019ApJ...874..171C, 2019ApJ...886L..13P, 2020ApJ...899...15M, 2023ApJ...947...74L}. In the full Zeeman sample, this selection excludes 23 blended components. Additionally, we do not include components detected toward 3C 98 and 3C 353 where artifacts and proximity to the GALFA-HI footprint edge preclude obtaining reliable emission maps. In the following analysis, we focus on the remaining 42 measurements.

\subsection{Statistical Methods}\label{section:statistical-methods}
Zeeman measurements of the 21-cm line are sensitive only to the line-of-sight projection of the total magnetic field, so we begin with a brief discussion of the statistical relationship between $B_{\text{LOS}}$ and $B_{\text{TOT}}$. In the following section, $B_{\text{LOS}}$ will denote the observed Zeeman measurement, and $B_{\mathrm{LOS,true}}$ will stand for the true underlying line-of-sight component of the total magnetic field strength. $B_{\text{LOS}}$ is derived from least-squares fitting of the Stokes $V$ spectrum and its probability follows the normal distribution,
\begin{equation}\label{eq:pdf-zee}
    P(B_{\text{LOS}} \mid B_{\mathrm{LOS,true}}) \sim \mathcal{N}\left(B_{\mathrm{LOS,true}}, \sigma(B_{\text{LOS}})\right),
\end{equation}
where $\sigma(B_{\text{LOS}})$ stands for the least-squares uncertainty. $B_{\text{LOS},\mathrm{true}}$ also carries a sign, positive for the field pointing away from the observer and vice versa. $B_{\mathrm{LOS,true}}$ depends on the total field strength $B_{\text{TOT}}$ and the inclination angle of the total magnetic field vector with respect to the observer $\gamma \in [0, \pi]$, and is given by $B_{\mathrm{LOS,true}}=B_{\text{TOT}} \cos{\gamma}$. When the field points toward or away from the observer, $B_{\mathrm{LOS,true}} = \pm B_{\text{TOT}}$, $\gamma = 0 \textrm{ or } \pi$, and when the field lies in the plane of the sky, $B_{\text{LOS},\mathrm{true}} = 0$, $\gamma = \pi/2$.
Assuming a random viewing angle, the probability distribution of the factor of $\cos{\gamma}$ follows $\mathrm{Uniform}(-1,1)$ \citep{Feller1991}, and we adopt the following parametrization
\begin{equation}\label{eq:pdf-blos}
   P\left(B_{\mathrm{LOS,true}} \mid B_{\text{TOT}}\right) \sim \mathrm{Uniform}\left(-B_{\text{TOT}} , B_{\text{TOT}}\right).
\end{equation}
The intrinsic distribution of the total magnetic field strength PDF($B_{\text{TOT}}$) in a population of CNM clouds is generally unknown, and might depend on other physical quantities, e.g., volume density. For a choice of PDF($B_{\text{TOT}}$) expressed in terms of model parameters $\alpha$, the likelihood function of a sample of $N$ independent Zeeman measurements is
\begin{equation}\label{eq:likelihood}
P\left(\mathrm{data} \mid \alpha\right)=\prod_n^{N} P(
B_{\text{LOS},n} \mid \alpha),
\end{equation}
where 
\begin{equation}
\begin{split}
& P(B_{\text{LOS},n} \mid \alpha) = \iint  P(B_{\text{LOS},n} \mid B_{\text{LOS},\mathrm{true},n}) \times \\
& P(B_{\text{LOS},\mathrm{true},n} \mid B_{\text{TOT},n}) \, P(B_{\text{TOT},n} \mid \alpha)\, \\
& \textrm{d}B_{\text{LOS},\mathrm{true},n} \,\textrm{d}B_{\text{TOT},n} .
\end{split}
\end{equation}

Common choices of the PDF of $B_{\text{TOT}}$ include the Dirac Delta distribution, 
\begin{equation}
P(B_{\mathrm{TOT}}|B_0) = \delta(B_{\mathrm{TOT}} - B_0),
\end{equation} describing a constant magnetic field strength across all clouds parametrized by $B_0$; a Uniform distribution 
\begin{equation}
P(B_{\mathrm{TOT}}|B_0) =
\begin{cases}
\frac{1}{B_0}, & \text{if } 0 \le B_{\mathrm{TOT}} \le B_0, \\
0, & \text{otherwise,}
\end{cases}
\end{equation}
where $B_{\text{TOT}}$ is distributed uniformly between 0 and a maximum value of $B_0$ \citep{2010ApJ...725..466C}; and a lognormal distribution described by
\begin{equation}
P(B_{\text{TOT}}|B_0,\sigma_0) = \frac{1}{B_{\text{TOT}}\sqrt{2\pi}\sigma_0}\; e^{-\left(\ln B_{\text{TOT}}-\ln B_0\right)^2/2\sigma_0^2},
\end{equation}
representing a Gaussian distribution of $\ln B_{\text{TOT}}$ with spread $\sigma_0$ \citep{2015MNRAS.451.4384T}. 

A specific choice of PDF$(B_{\text{TOT}})$ together with the likelihood in Equation \ref{eq:likelihood} defines a simple model for a sample of independent line-of-sight field strength measurements. In practice, least-squares uncertainties on measurements toward the same background source and at similar velocities might be correlated, but existing datasets do not contain information that would allow us to model such correlation. We use random draws from this model in null hypothesis testing in the subsequent analysis. \cite{2010ApJ...725..466C} found that the Millennium sample does not favor any particular model of PDF$(B_{\text{TOT}})$. Consistent with this finding, our null hypothesis tests are not highly sensitive to the assumed form of PDF$(B_{\text{TOT}})$.

\subsection{Correlation between $|B_{\text{LOS}}|$ and HI Filament Disorder}\label{sec:correlation}

\begin{figure}[ht]
        \centering
        \includegraphics[trim={1.5cm 0.1 1cm 0cm}, scale=0.6]{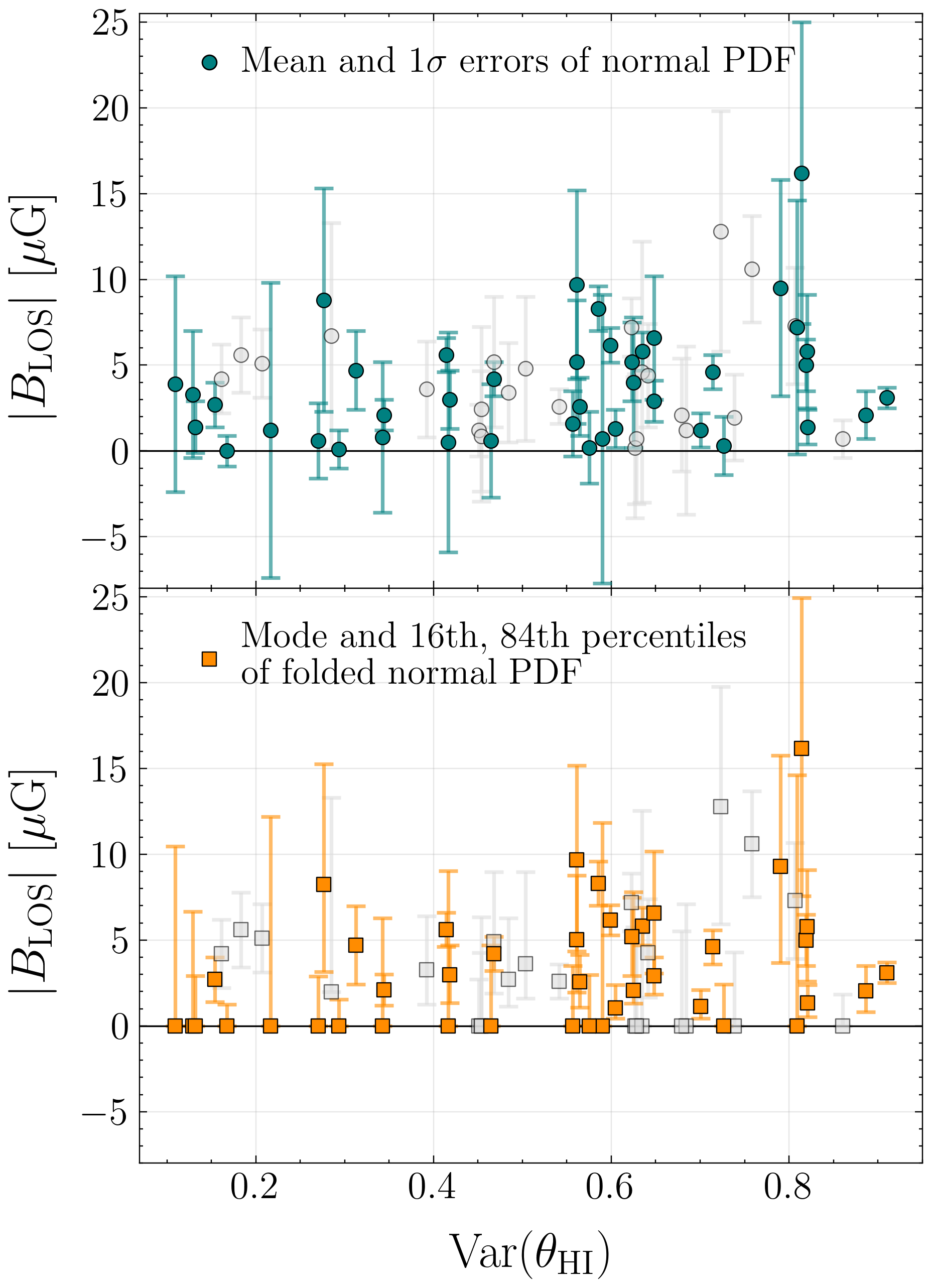}
        \caption{Zeeman measurements plotted against the circular variance of HI filament orientations $\text{Var}(\theta_{\mathrm{RHT}})$ computed in a $d_{\mathrm{POS}} < 2.5^{\circ}$ region around each radio source. Top: We plot the mean and $1 \sigma$ errors of a normal distribution centered at $|B_{\text{LOS}}|$ with uncertainties of the directional $B_{\text{LOS}}$. Gray points represent measurements we classified as blended in \S\ref{section:crowding}. Bottom: Orange points represent the mode of a folded normal distribution of $|B_{\text{LOS}}|$. The error bars represent the $16$th and $84$th percentiles of the folded normal PDF.
        }
        \label{fig:Blos-vs-disorder}
    \end{figure}

We examine the hypothesis that the dispersion of plane-of-sky filament orientations is correlated with the line-of-sight magnetic field strength. In Figure \ref{fig:Blos-vs-disorder}, we plot the magnitude of Zeeman measurements $|B_{\text{LOS}}|$ against the variance of HI filament orientation angles $\mathrm{Var}(\theta_{\mathrm{HI}})$. Filament dispersion is computed in a circular patch of diameter $d_{\text{POS}}=2.5^{\circ}$ centered at the background radio source and the velocity of the corresponding absorption component. $|B_{\text{LOS}}|$ and HI filament disorder show a weak positive correlation (Spearman $\rho = 0.31$). In the top panel of Figure \ref{fig:Blos-vs-disorder}, we follow previous authors in plotting $|B_{\text{LOS}}|$ with $1\sigma$ Gaussian errors of $B_{\text{LOS}}$. Since $|B_{\text{LOS}}|$ follows a folded-normal distribution \citep{2025MNRAS.540.2762W}, in the bottom panel of Figure \ref{fig:Blos-vs-disorder} we visualize the data using the mode and the 16th, 84th percentiles of the folded-normal distribution.

The statistical significance of the $|B_{\text{LOS}}|$ -  $\mathrm{Var}(\theta_{\mathrm{HI}})$ correlation deserves careful consideration due to the small size of our sample and the fact that Zeeman measurement uncertainties are heteroskedastic. To evaluate whether the observed correlation could arise randomly, we simulate null-hypothesis samples by drawing $B_{\text{LOS,true}}$ values from the model described in \S\ref{section:statistical-methods}, adopting PDF($B_{\text{TOT}}$) representative of the observed Millennium measurements and preserving the Zeeman measurement uncertainty associated with each $\mathrm{Var}(\theta_{\mathrm{HI}})$.

To find the parameters of PDF($B_{\text{TOT}}$) that best fit the dataset, we first maximize the likelihood defined in Equation \ref{eq:likelihood} with respect to the sample of 42 non-blended Zeeman measurements. We obtain $B_0 = 6.2 \: \mu$G and $B_0 = 9.6 \: \mu$G for the Delta and Uniform PDFs respectively, in agreement with the Bayesian model of \cite{2010ApJ...725..466C}, and $B_0 = 1.7, \sigma = 0.2$ for the log-normal model. With these PDF($B_{\text{TOT}}$), we use the model defined in \S\ref{section:statistical-methods} to draw 42 random $B_{\text{LOS}}$, adopting the errors $\sigma(B_{\text{LOS}})$ reported by \citetalias{2004ApJS..151..271H} in Equation \ref{eq:pdf-blos}. We then calculate the Spearman $\rho$ between the absolute values of the simulated measurements and the observed $\mathrm{Var}(\theta_{\mathrm{HI}})$ while preserving the pairings between $\sigma(B_{\text{LOS}})$ and $\mathrm{Var}(\theta_{\mathrm{HI}})$. We generate $10^6$ such samples and find a Spearman $\rho$ at least as positive as in the Millennium data in only 1.4\%, 1.0\%, 1.3\% of cases for models with Delta, Uniform and log-normal PDF($B_{\text{TOT}}$), respectively. This level of statistical significance suggests that the observed correlation is unlikely to have arisen due to random chance and motivates our further investigation of the trend. 

We find a weaker ($\rho = 0.21$, $p=0.04$) correlation using all 62 Zeeman measurements with usable HI emission maps, including the components we classified as blended. Conversely, we find a stronger ($\rho = 0.40$, $p=0.004$) correlation when we increase the blending threshold to 0.80, including only 38 measurements. We compare the samples of blended and non-blended components using the Kolmogorov-Smirnov (K-S) test and find no significant difference in the distributions of $|B_{\text{LOS}}|$, while the distributions of $\sigma(B_{\text{LOS}})$ differ, with the non-blended sample containing generally lower $\sigma(B_{\text{LOS}})$. The difference in Zeeman measurement errors between the two samples is expected, as absorption components that dominate the CNM emission typically have higher optical depth, resulting in a more sensitive Zeeman measurement.  Distributions of other physical parameters ($T_s$, $N(\mathrm{HI})$, $\sigma_{v, \textrm{NT}}$) do not differ significantly between the two samples. The enhanced $|B_{\text{LOS}}|$--$\mathrm{Var}(\theta_{\mathrm{HI}})$ correlation in the non-blended sample suggests that either our selection identifies Zeeman measurements with a stronger connection to the observed HI emission morphology, or that the selected components provide more reliable Zeeman measurements due to lower uncertainties and reduced degeneracies in the least-squares fit.

\subsection{Angular Scale and Algorithmic Dependence}

\begin{figure}[ht]

    \centering
    \includegraphics[trim={0.5cm 0cm 0cm 0cm}, scale=0.97]{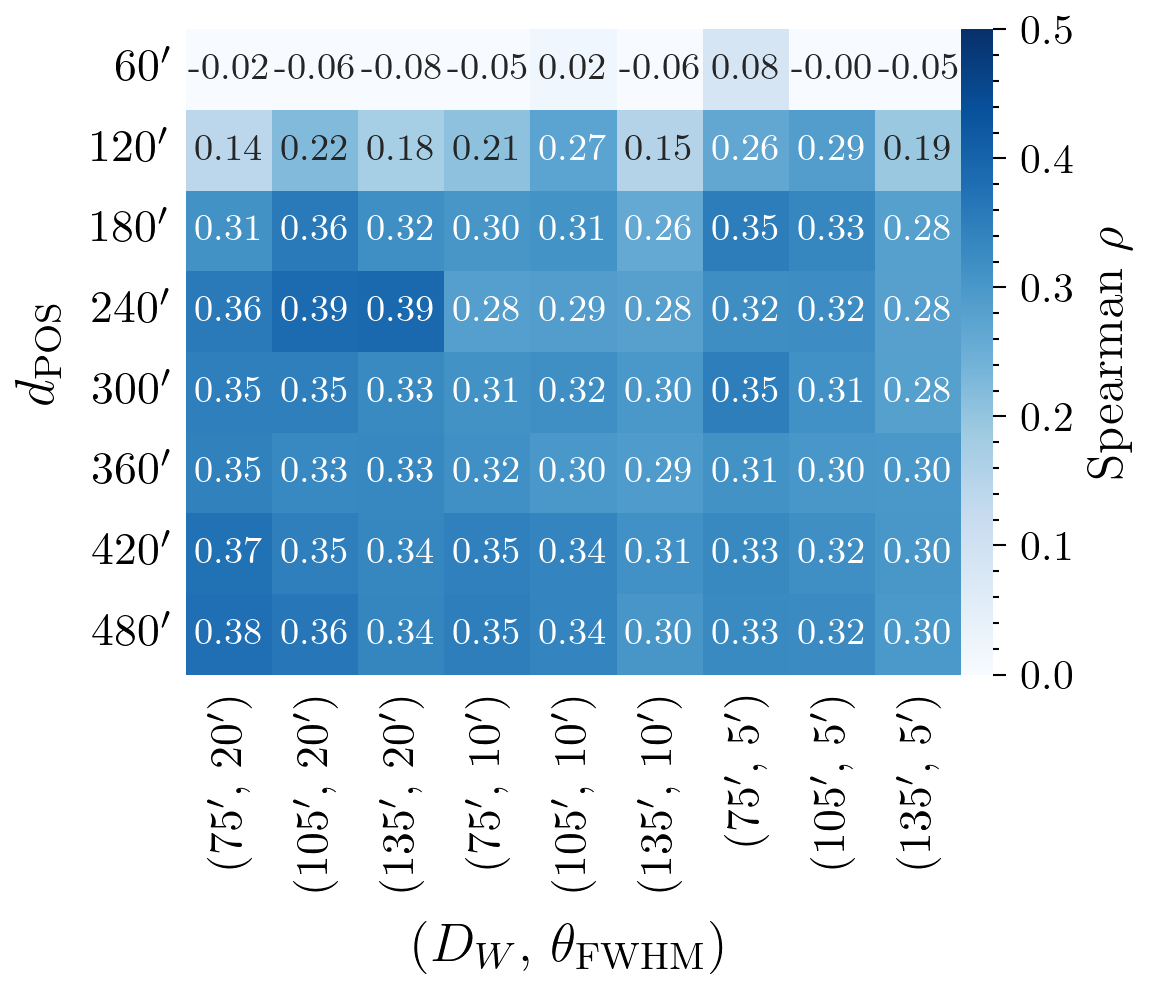}
    \caption{Spearman correlation between $|B_{\text{LOS}}|$ and $\mathrm{Var}(\theta_{\mathrm{HI}})$ as a function of the plane-of-sky diameter $d_{\text{POS}}$ of the region where $\mathrm{Var}(\theta_{\mathrm{HI}})$ is computed and as a function of RHT parameters: the window length $D_W$ and the smoothing radius $\theta_{\mathrm{FWHM}}$, at a fixed RHT threshold $Z_{\text{RHT}}=0.75$. We find that the result is not strongly sensitive to the choice of RHT parameters. The correlation is not present at scales smaller than the RHT window length, but persists to large angular scales.}
    \label{fig:scale}
\end{figure} 

We initially ran the RHT with $D_W=105'$ and $\theta_{\mathrm{FWHM}}=5'$ following the results of \cite{2023ApJ...945...72A}, who used Planck maps and the HI4PI survey \citep{2016A&A...594A.116H} to demonstrate that the HI emission structures that best correlate with polarized dust emission are long ($D_W = 80'-160'$) and thin ($\theta_{\mathrm{FWHM}}$ similar to the HI4PI beam size). To investigate the sensitivity of the observed correlation to the choice of RHT parameters, we test every combination of $D_W = [75', 105', 135']$ and $\theta_{\mathrm{FWHM}} = [5', 10', 20']$. We summarize the impact of the RHT parameters on the observed correlation in Figure \ref{fig:scale}, and conclude that the correlation is robust to parameter choice. We also tested a lower RHT detection threshold $Z_{\text{RHT}}=0.25$, which retains structures that are less filamentary, and found no $|B_{\text{LOS}}|$--$\mathrm{Var}(\theta_{\mathrm{HI}})$ correlation, suggesting that the observed association relies on features that display distinctly coherent, filamentary morphology. We also find the observed trends robust to increasing velocity channel width from $1.7 \,\textrm{km}\,\textrm{s}^{-1}$ to $3 \,\textrm{km}\,\textrm{s}^{-1}$, which is expected given that both channel widths are comparable to typical linewidths of our Zeeman components. 

The $|B_{\text{LOS}}|$--$\mathrm{Var}(\theta_{\mathrm{HI}})$ correlation varies with the plane-of-sky scale over which HI filament disorder is computed, as shown in Figure \ref{fig:scale}. No significant correlation is found at $d_{\text{POS}} < 2^{\circ}$ for any choice of RHT parameters. At scales below $d_{\text{POS}} \approx 2^{\circ}$, which are comparable to the RHT window length, only a few distinct structures are detected per region, and $\mathrm{Var}(\theta_{\mathrm{HI}})$ is not robust due to insufficient filament statistics: e.g., a single filament is not informative of filament dispersion. The correlation does not significantly decrease even at $d_{\text{POS}} = 8^{\circ}$, the largest angular scale we have tested. This is possibly because the orientation of HI filaments remains correlated on these scales, as found by Putman et al. (2025, in prep), who analyzed properties of discrete HI filaments across the whole GALFA-HI footprint. We also note that above $d_{\text{POS}} \approx 5^{\circ}$, there is significant overlap in the GALFA-HI maps between some adjacent sight lines.

\begin{figure*}[ht]
    \centering
    \includegraphics[trim={0cm 0cm 0cm 0cm}, scale=0.5]{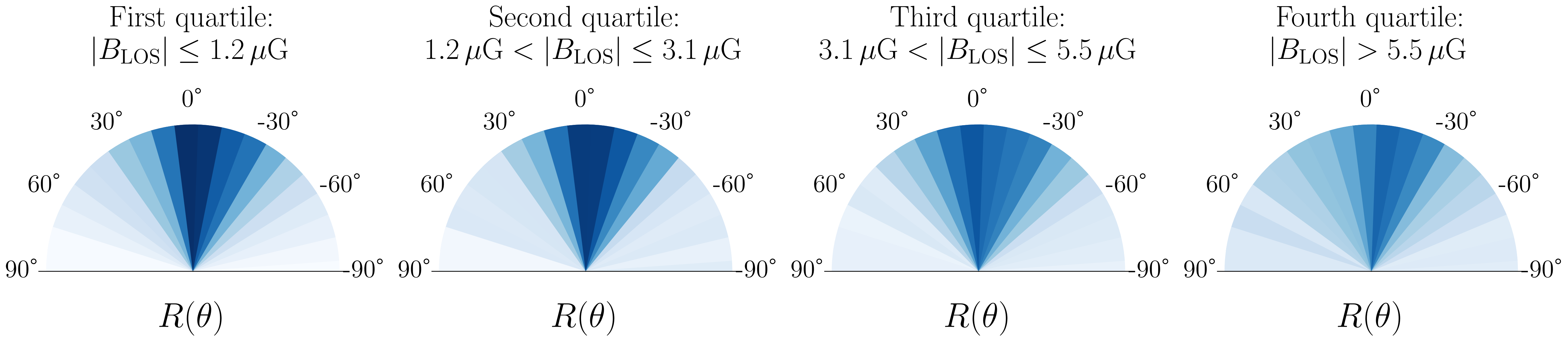}
    \caption{Spatially averaged RHT output $R(\theta)$ centered around the mean orientation and stacked in quartiles of Zeeman field strength. HI emission maps corresponding to lower Zeeman field strengths show more sharply peaked distributions of filament orientations (dark blue), and vice versa.}
    \label{fig:stacked-r-theta}
\end{figure*} 

Until now, we have quantified the dispersion of HI filaments using the circular variance of orientation angles $\theta_{\text{HI}}$. If the $|B_{\text{LOS}}|$--$\mathrm{Var}(\theta_{\mathrm{HI}})$ correlation is robust, it should be independent of the choice of summary statistic that quantifies HI filament dispersion. Here we explore an alternative measure of plane-of-sky filament disorder, the normalized RHT output function $R(x,y,\theta)$, representing the probability distribution of orientations of linear structures passing through each pixel $(x,y)$. For each of our GALFA-HI cutouts, we compute the spatial mean of $R(x,y,\theta)$ within $d_{\text{POS}}<2.5^{\circ}$ and center the resulting $R(\theta)$ distribution around its mean angle. In Figure \ref{fig:stacked-r-theta}, we show the stacked, normalized $R(\theta)$ signal in quartiles of the Zeeman line-of-sight field strength. Maps with higher $|B_{\text{LOS}}|$ have wider $R(\theta)$ distributions, in agreement with the result we obtained by quantifying filament disorder using $\text{Var}(\theta_{\mathrm{RHT}})$. 

We verify that the $|B_{\text{LOS}}|$--$\mathrm{Var}(\theta_{\mathrm{HI}})$ correlation does not arise due to an association between $|B_{\text{LOS}}|$ and another morphological property of the HI emission maps. At $D_W=105'$ and $\theta_{\mathrm{FWHM}}=5'$, we find no correlation between $|B_{\text{LOS}}|$ and the spatial sum of RHT intensity $I_{\mathrm{RHT}}$. This suggests an association of Zeeman measurements with the relative alignment of filaments rather than with the prominence of filamentary structures in a map. We also verify the lack of correlation between $|B_{\text{LOS}}|$ and the number of pixels with non-zero RHT output, which could bias $\mathrm{Var}(\theta_{\mathrm{HI}})$ toward lower values in regions where filamentary HI structures are sparse. As an additional test, we compute $\mathrm{Var}(\theta_{\mathrm{HI}})$ in regions of equal RHT intensity or an equal number of RHT-activated pixels rather than at fixed angular scale $d_{\text{POS}}$ and find no decrease in the original correlation. We conclude that the observed trend is particularly linked to the relative alignment of HI filaments, while measures reflecting the covering fraction or prominence of filaments fail to show a similar trend.

\subsection{Dependence on Velocity and Position}

We perform several statistical tests to determine whether the observed correlation is sensitive to the exact velocity or position at which we compute the dispersion of HI filaments. 

To test sensitivity to the plane-of-sky position, for each Zeeman component we compute $\mathrm{Var}(\theta_{\mathrm{HI}})$ in 4 non-overlapping $d_{\text{POS}}=2.5^{\circ}$ regions around positions that are offset from the radio source by $255'$. The number and displacement of offset positions were chosen so that their boundaries are separated by at least one RHT window length $D_W$ from the initial position, ensuring that the estimates of filament orientation dispersions are independent. We then generate $10^6$ random samples of 42 $\mathrm{Var}(\theta_{\mathrm{HI}})$, where for each Zeeman component, $\mathrm{Var}(\theta_{\mathrm{HI}})$ is chosen randomly from one of the four displaced positions, and calculate the Spearman $\rho (|B_{\text{LOS}}|, \mathrm{Var}(\theta_{\mathrm{HI}}))$ correlation coefficient for each sample, finding $\rho$ at least as positive as the observed one in $7\%$ of the samples.

We next test whether the observed trends are sensitive to the velocity at which $\mathrm{Var}(\theta_{\mathrm{HI}})$ is computed. For each Zeeman component at velocity $v_0$, we extract GALFA-HI maps centered at $v_0 + \delta v$ and $v_0 - \delta v$ where $\delta v = 6 \,\textrm{km}\,\textrm{s}^{-1}$, comparable to the maximum linewidth of a Zeeman component in our sample. At these displaced velocities, we find a decrease in the $|B_{\text{LOS}}|$--$\mathrm{Var}(\theta_{\mathrm{HI}})$ correlation down to $\rho \approx 0.15-0.20$. When we repeat the test with a larger velocity displacement of $\delta v = 9 \,\textrm{km}\,\textrm{s}^{-1}$, the correlation disappears entirely. A limitation of this procedure is that some displaced channels do not contain high signal-to-noise HI emission.

In summary, shifts in plane-of-the-sky position and velocity reduce the observed correlation but do not always eliminate it, implying that the association between $\mathrm{Var}(\theta_{\mathrm{HI}})$ and $B_{\text{LOS}}$ may reflect large-scale variations in HI morphology and line-of-sight magnetic field strength across the sky, rather than local conditions near the Zeeman absorber. Due to the unbiased target selection of the Millennium Survey, the lines of sight with Zeeman measurements probe a variety of environments, from high-latitude diffuse gas to the outskirts of molecular clouds in the Gould Belt, as shown in Figure \ref{fig:zeeman-oh-co}. In the next section, we use multi-wavelength datasets to explore the possibility that the $|B_{\text{LOS}}|$--$\mathrm{Var}(\theta_{\mathrm{HI}})$ correlation arises due to differences in the Galactic environment.

\section{Environment Characterization}\label{sec:environment}
To better characterize the Galactic environments probed by the Millennium Zeeman sight lines, we utilize dust emission maps and spectral lines of CO and OH. In this analysis, which does not rely on the morphological quantities we investigated in the previous section, we use all 66 Millennium measurements.

\subsection{Variations with Column Density}

\begin{figure*}{}
    \centering
    \includegraphics[trim={0cm 0cm 0cm 0cm}, clip, scale=0.72]{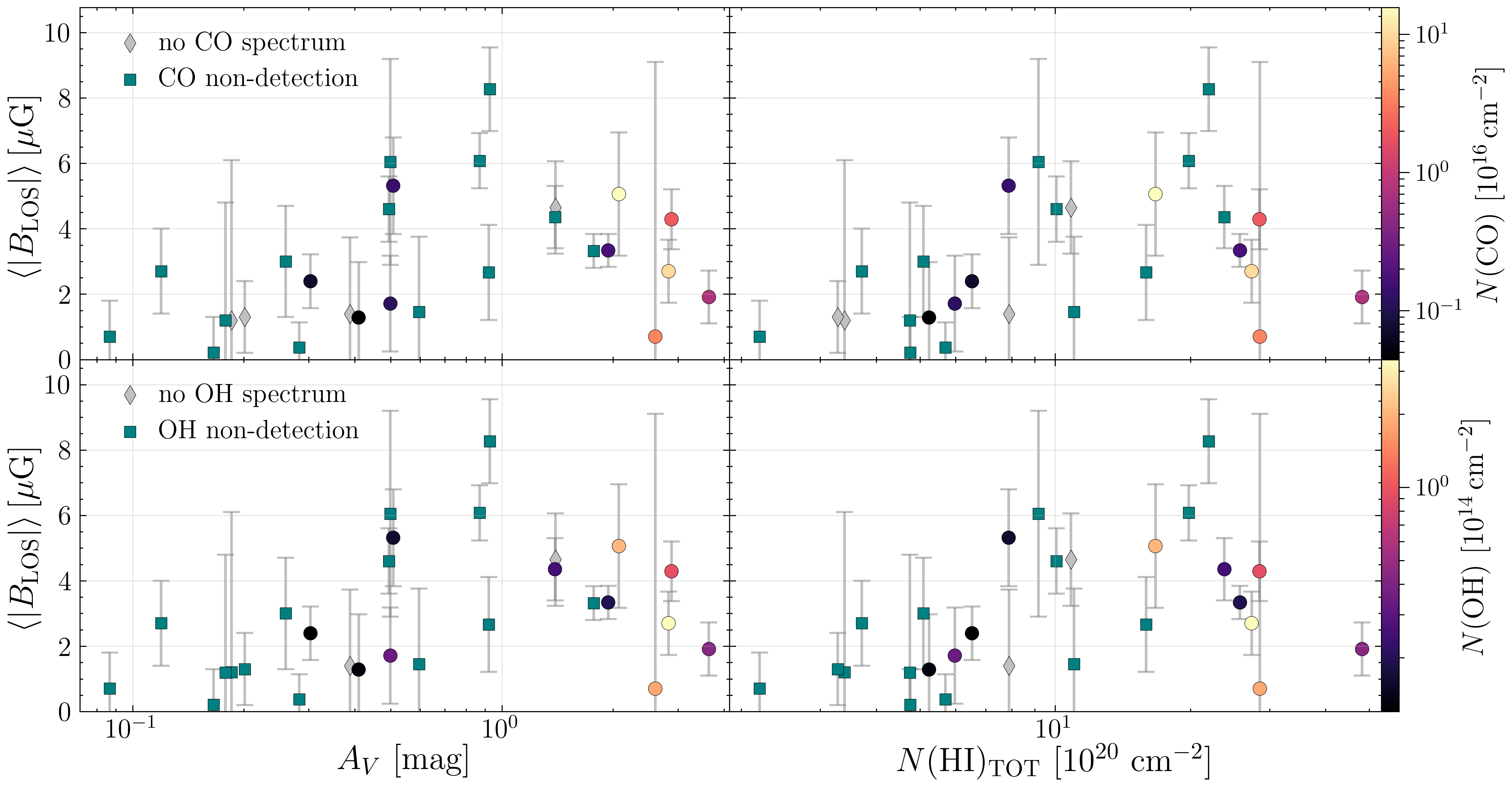}
    \caption{Inverse variance-weighted mean $|B_{\text{LOS}}|$ along the line of sight plotted against integrated density tracers: dust visual extinction $A_V$ (left) and the total HI column density (right). In the top row, points are color-coded by the integrated column density of CO; the bottom row shows the same but for OH, derived based on the 1667 Hz line. For each molecular tracer, teal squares mark non-detections while gray diamonds mark targets with no molecular line spectrum in \cite{2018ApJS..235....1L}.}
    \label{fig:sightline-blos-AV}
\end{figure*}

In Figure \ref{fig:sightline-blos-AV}, we plot the average Zeeman-inferred $|B_{\text{LOS}}|$ per sight line against the Planck-derived dust extinction $A_V$ (left column) and the total atomic hydrogen column density $N$(HI) derived from the Millennium spectra (right column). The mean $|B_{\text{LOS}}|$ for each sight line is computed by averaging the absolute values of Zeeman measurements, weighted by their inverse variance. We find that $|B_{\text{LOS}}|$ tends to be larger in sight lines with higher dust extinction. A Spearman correlation yields $\rho (A_V, \langle|B_{\text{LOS}}|\rangle) = 0.48$. To assess its significance, we employ the same strategy as in the previous section: we generate random samples of individual Zeeman measurements from a representative distribution of PDF($B_{\text{TOT}}$) while preserving, for each $A_V$ measurement, the number of Zeeman components along the line of sight and their uncertainties. Recomputing $\rho (A_V, \langle|B_{\text{LOS}}|\rangle)$ for each random sample confirms the statistical significance of the observed trend ($p \approx 0.005$ across all models); the results are similar for $\rho (N\text{(HI)}, \langle|B_{\text{LOS}}|\rangle)$. Notably, we observe a similar relationship in both columns of Figure \ref{fig:sightline-blos-AV}, despite $N$(HI) tracing only atomic gas and $A_V$ tracing both atomic and molecular gas. 

We interpret Figure \ref{fig:sightline-blos-AV} as tentative evidence of a connection between Zeeman measurements and the larger Galactic environment, either due to global variations in $B_{\text{TOT}}$ or inclination angle. A population of clouds lying along the line of sight with random viewing angles would have a higher mean $|B_{\text{LOS}}|$ if their underlying $B_{\text{TOT}}$ were larger on average. Alternatively, a population of clouds lying along a line of sight with magnetic field strengths drawn from a random distribution of  $B_{\text{TOT}}$ could have a higher or lower mean $|B_{\text{LOS}}|$ if the large-scale magnetic field in that direction has a preferred orientation with respect to the line of sight. For example, sight lines with lower $A_V$, located primarily at high Galactic latitudes, could have magnetic fields that are more tangential to the observer and therefore have lower $|B_{\text{LOS}}|$. Either interpretation points to systematic differences in mean $B_{\text{LOS}}$ across the sight lines probed by the Millennium Survey.

\subsection{Dependence on the Presence of Molecular Gas Tracers}

OH and CO molecules are indirect tracers of molecular hydrogen (H$_2$), which lacks rotational emission lines at the low temperatures typical of molecular clouds \citep{1970ApJ...161L..43W, 2018ApJ...862...49N}. HI is found around molecular clouds, as it provides the raw material for H$_2$ formation and is mixed with dust, which shields H$_2$ from dissociation by UV photons \citep{2012ApJ...748...75L}. We use OH and CO molecules to distinguish lines of sight containing primarily atomic gas from lines of sight where HI Zeeman measurements might be probing the surroundings of molecular clouds.

To investigate whether higher average magnetic field strength can be associated with molecular environments, we color-code points in Figure \ref{fig:sightline-blos-AV} by the line-of-sight integrated CO and OH column densities reported by \cite{2018ApJS..235....1L}. At the lowest $N$(HI), CO and OH are not detected, while the sight lines with highest $N$(HI) have significant column densities either in OH, CO, or both tracers. We categorize sight lines as molecular if an OH or CO detection exists at any velocity in the \cite{2018ApJS..235....1L} spectra, or if there is CO emission at any velocity in the \cite{2022ApJS..262....5D} spectra. The remaining sight lines are categorized as atomic. No systematic difference in the distribution of individual $|B_{\text{LOS}}|$ measurements or inverse variance-weighted sight line means $\langle|B_{\text{LOS}}|\rangle$ is found between molecular and atomic sight lines using a K-S test ($p=0.97$, $p=0.3$ for $|B_{\text{LOS}}|$, $\langle|B_{\text{LOS}}|\rangle$, respectively).

\begin{figure}{}
    \centering
    \includegraphics[trim={0.cm 0cm 0cm 0cm}, clip, scale=0.58]{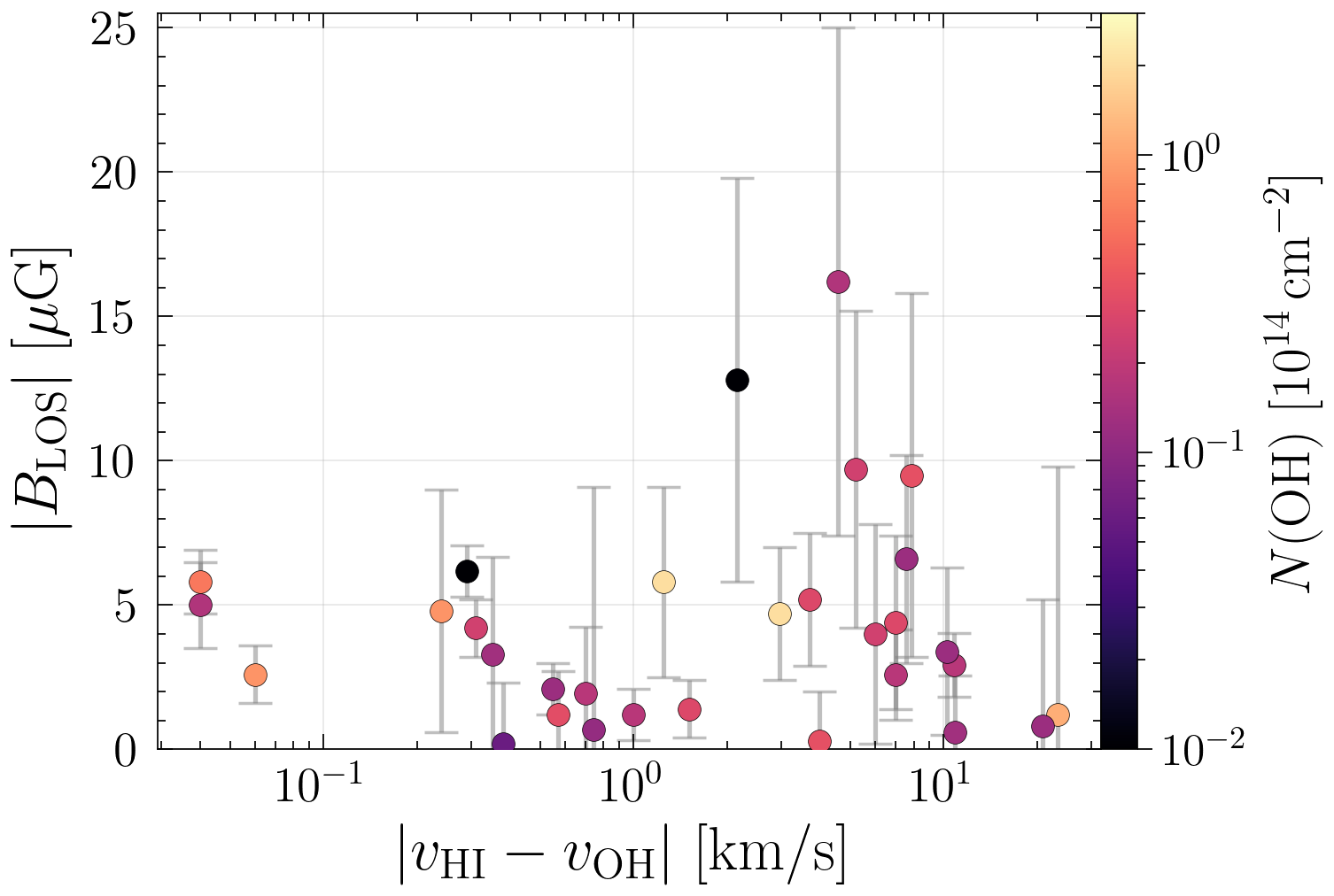}
    \caption{Zeeman field strength measured in HI absorption as a function of velocity offset between HI absorption component and the nearest OH 1667 Hz absorption component, for sight lines where OH is detected. }
    \label{fig:velHIOH}
\end{figure}

A similar analysis on the level of individual velocity components is less straightforward. While HI, OH, and CO measurements are all spectral, line components at the same velocity might not be physically associated across tracers, and conversely, distinct velocity components might be co-spatial \citep{2013ApJ...777..173B}. For instance, \cite{2019A&A...622A.166S} compared the morphology of CO and HI emission in the Galactic plane using Histograms of Oriented Gradients and found that some regions show spatial correlations between the two tracers in channels separated by a few kilometers per second. We have attempted to morphologically correlate CO and HI emission in Millennium sight lines, but were limited by the insufficient resolution of CO maps of \cite{2022ApJS..262....5D} away from the Galactic plane where most Zeeman sight lines lie. Keeping in mind that velocity is not a direct probe of distance, in Figure \ref{fig:velHIOH}, we plot the magnetic field strength as a function of the velocity difference between the HI and OH components in the sight lines that have an OH detection. Every possible pair of OH and HI components identified in a spectrum is shown. We observe that HI components that are co-spectral with OH absorbers do not exhibit preferentially higher magnetic field strengths, but measurements with the highest signal-to-noise ratios are within $1 \,\textrm{km}\,\textrm{s}^{-1}$ of OH lines. The results are qualitatively similar when CO components are considered in place of OH components. 

We compare the Zeeman magnetic field strengths between HI and OH components for the 4 Millennium sight lines that have OH Zeeman measurements reported by \cite{2019ApJ...884...49T}. In these 4 sight lines, the only $5 \sigma$ Zeeman detection in HI absorption occurs toward 3C 133 and agrees with the OH result in magnitude, with $B_{\text{LOS}}(v_{\mathrm{OH}}=7.7 \,\textrm{km}\,\textrm{s}^{-1}) = -5.9 \pm 1.8~\mu \textrm{G}$ and $B_{\text{LOS}}(v_{\mathrm{HI}}=8.0 \,\textrm{km}\,\textrm{s}^{-1}) = 5.8 \pm 1.1~\mu \textrm{G}$, but not in direction. This sign difference is difficult to understand and may be due to a sign error in one of the two published results. More data are needed to infer whether Zeeman measurements in HI and OH absorption lines are universally correlated due to the low signal-to-noise ratio and small target overlap in the available samples. 

\section{Scale and Structure}\label{sec:structure}
This section uses all available Millennium measurements and 3D dust maps to examine the scale and structure probed by Zeeman measurements in HI absorption. 

\subsection{Spatial and Velocity Coherence}

\begin{figure*}{}
    \centering
    \includegraphics[trim={0cm 0cm 0cm 0cm}, clip, scale=0.9]{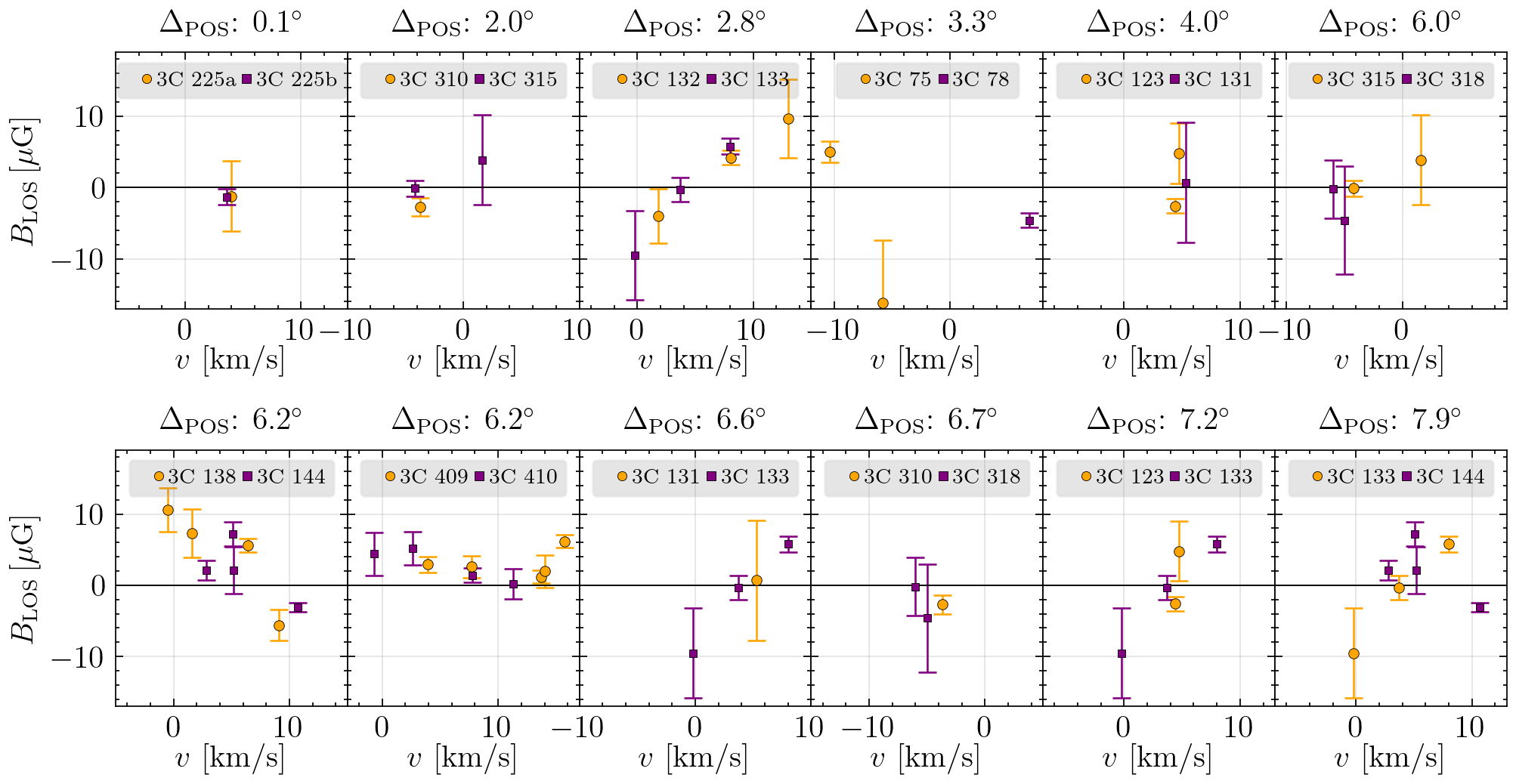}
    \caption{Zeeman measurements toward background sources with the smallest angular separations in the survey. Each panel shows Zeeman measurements toward two sight lines, and panels are ordered by the angular separation of the background sources. We find a general agreement, both in strength and direction, between $B_{\text{LOS}}$ at similar velocities. The first $> 3 \sigma$ disagreement between co-spectral ($|v_i-v_j|<3\,\textrm{km}\,\textrm{s}^{-1}$) components is observed toward targets separated by $7.9^{\circ}$ (3C 133 and 3C 144 (Tau A)). Each plot spans $20\,\textrm{km}\,\textrm{s}^{-1}$ but the central velocity varies. }
    \label{fig:sky-neighbors}
\end{figure*}

We investigate the coherence of the magnetic field traced by Zeeman splitting. In Figure \ref{fig:sky-neighbors}, we examine the structure in Zeeman measurements on the plane of the sky. There are 12 pairs of radio sources in the sample with an angular separation $\leq 8^{\circ}$. All pairs of HI absorption components with angular separation $\leq 7.5^{\circ}$ and velocity separation $< 3\,\textrm{km}\,\textrm{s}^{-1}$ agree in both magnetic field strength and direction within $1.5 \sigma$. The first $3\sigma$ difference between co-spectral components is observed at an angular separation of $7.9^{\circ}$, between the $3.7 \,\textrm{km}\,\textrm{s}^{-1}$ component of 3C 144 (Tau A) and the $5.1 \,\textrm{km}\,\textrm{s}^{-1}$ component of 3C 133. 

\begin{figure}{}
    \centering
    \includegraphics[trim={0cm 0cm 0cm 0cm}, clip, scale=0.69]{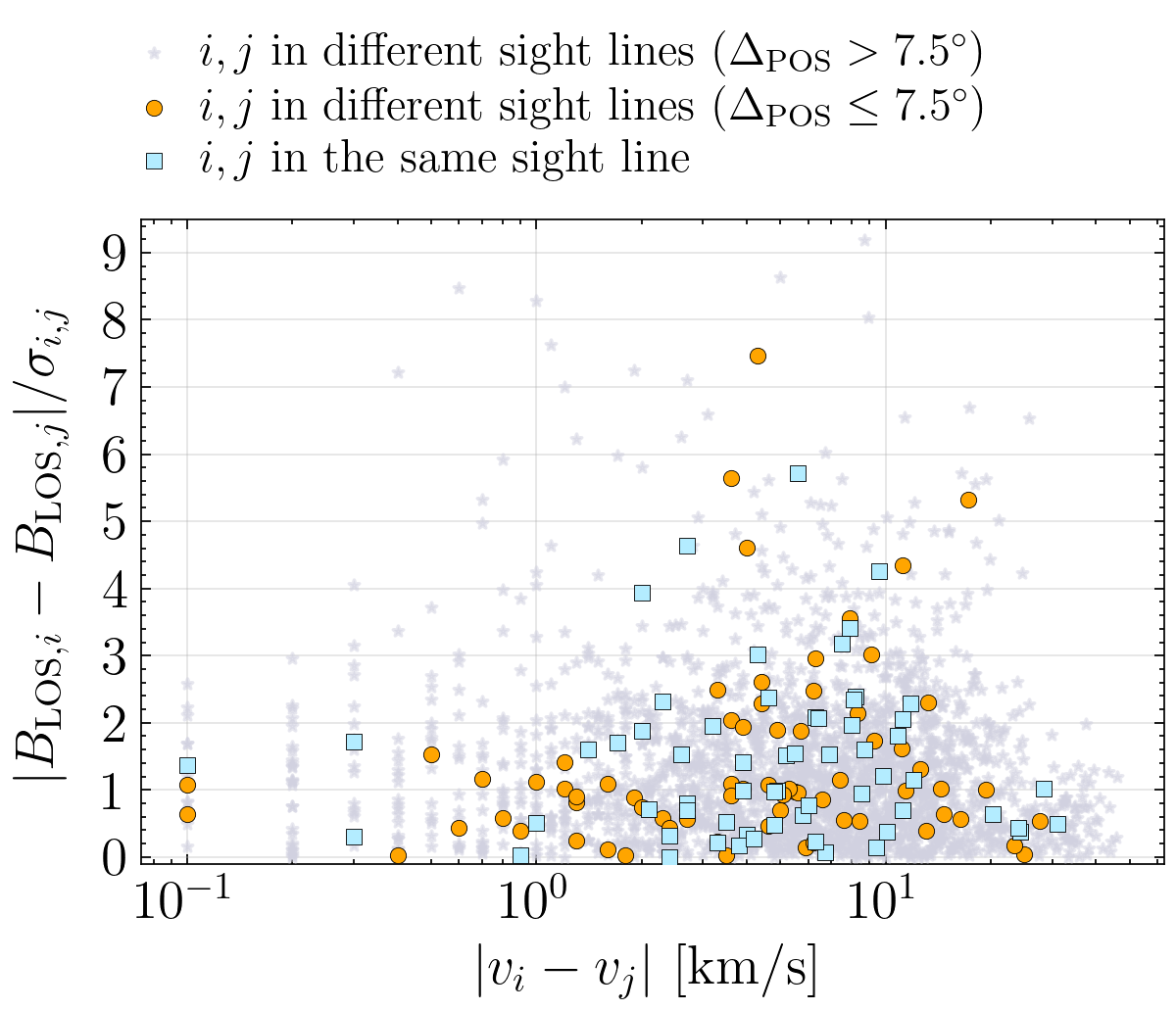}
    \caption{For each component pair $i, j$ in the Millennium sample, we plot the statistical significance of the Zeeman measurement difference, $|B_{\text{LOS},i} - B_{\text{LOS},j}| / \sigma_{i,j}$, as a function of their radial velocity separation. We show pairs that lie along the same line of sight in blue squares, different lines of sight with a large angular separation ($>7.5^{\circ}$) in gray stars, or different lines of sight with a small ($\leq 7.5^{\circ}$) angular separation in in orange circles.}
    \label{fig:vel-neighbors}
\end{figure}

To examine the coherence of Zeeman measurements in position and velocity space, for each pair $(i,j)$ of Zeeman measurements in the sample, we compute the statistical significance of their difference, $|B_{\text{LOS},i} - B_{\text{LOS},j}|/\sigma_{i,j}$, where we take into account the field direction, and where $\sigma_{i,j} = (\sigma(B_{\text{LOS},i})^2 + \sigma(B_{\text{LOS},j})^2)^{1/2}$. In Figure \ref{fig:vel-neighbors}, we plot $|B_{\text{LOS},i} - B_{\text{LOS},j}|/\sigma_{i,j}$ as a function of the velocity difference for pairs of Zeeman measurements belonging to three samples: pairs that lie along the same line of sight, different lines of sight with a large angular separation ($>7.5^{\circ}$), or different lines of sight with a small ($\leq 7.5^{\circ}$) angular separation. The choice of $7.5^{\circ}$ for the angular separation split is motivated by our findings in Figure \ref{fig:sky-neighbors}. While most data pairs agree within 3$\sigma$ across the three samples due to the low signal-to-noise ratio of Zeeman measurements, at low-velocity separations, pairs of points that are either in the same sight line or in closely separated sight lines show a qualitatively higher degree of coherence than pairs of points between sight lines that are far apart. 

Considering only co-spectral components with $|v_{i}-v_{j}|< 3\,\textrm{km}\,\textrm{s}^{-1}$, we compare $|B_{\text{LOS},i} - B_{\text{LOS},j}|/\sigma_{i,j}$ in the three samples using a K-S test and find that pairs of measurements toward targets with a small angular separation (orange circles) and targets with a large angular separation (gray stars) follow different distributions ($p=0.002$). The result is not highly sensitive to the exact velocity difference used to define co-spectral components. Similarly, we compare the co-spectral pairs along the same line of sight (blue squares) and toward targets with large angular separation (gray stars) and find weaker but significant evidence that they come from different distributions ($p=0.02$, but the result is more sensitive to the choice of the velocity threshold). Because the quantities $|B_{\text{LOS},i} - B_{\text{LOS},j}|/\sigma_{i,j}$ are pairwise combinations of Zeeman measurements, some share a common measurement, resulting in a non-zero covariance. We implement an additional test to validate the statistical significance of the KS test. We use the random samples of Zeeman measurements drawn from a representative PDF($B_{\text{TOT}}$) introduced in \S\ref{section:statistical-methods}), which preserve the structure of the sample in $\sigma({B_{\text{LOS}}})$, velocity separations, and angular separations between targets. We then measure the fraction of samples in which the K-S statistic is at least as large as in the real dataset and find consistent results ($p=0.004$ and $p=0.03$ for the two pairs of subsamples). We conclude that there is enhanced coherence in magnetic field strength at small angular ($\lesssim 7.5^\circ$) and velocity separations ($\lesssim 3\,\textrm{km}\,\textrm{s}^{-1}$). While we expect a similar degree of coherence in co-spectral components along the same line of sight, the signal might be weaker due to the smaller sample size, because velocity does not directly map to distance, or because each pair of measurements is obtained from the same spectrum and can have a significant overlap in velocity line profiles, leading to correlated errors. In \S\ref{sec:blos-HI-correlation}, we labeled many of the co-spectral components toward the same target as blended in emission and determined they correlate less with HI filament disorder.

\subsection{3D Dust Profiles toward Millennium Targets}

\begin{figure*}{}
\centering
\includegraphics[trim={0cm 0.15cm 0cm 0.0cm}, clip, scale=0.8]{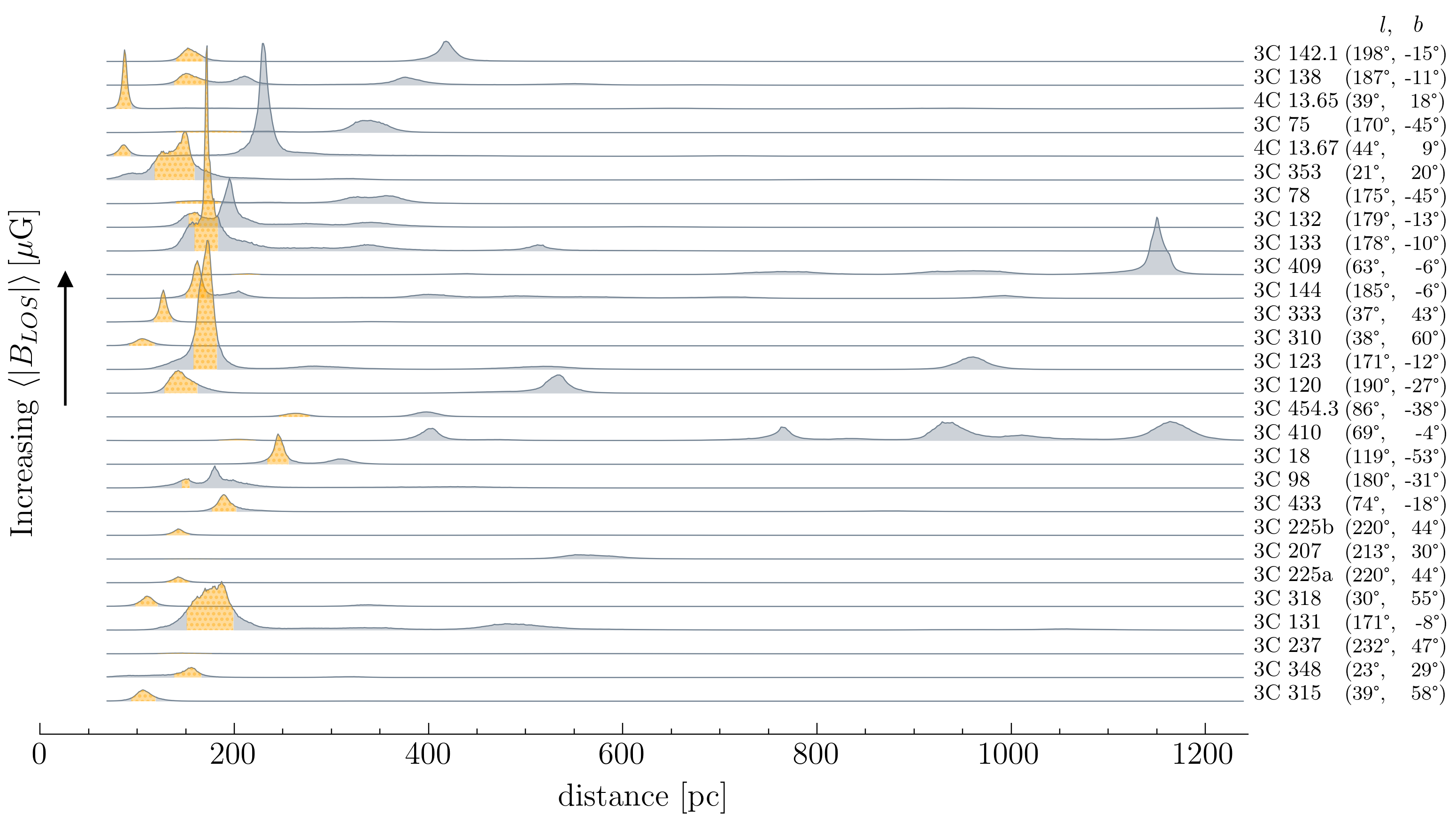}
\caption{Dust extinction per parsec toward Millennium Zeeman targets extracted from \cite{2024A&A...685A..82E} 3D dust maps. The dust profiles have been averaged over the 12 posterior samples. In dotted orange, we marked the boundaries of the Local Bubble from the model of \cite{2024ApJ...973..136O}. The sight lines are sorted in the increasing order of $\langle|B_{\text{LOS}}|\rangle$, the inverse-variance weighted mean of $|B_{\text{LOS}}|$.}
\label{fig:dust-profiles}
\end{figure*}

In Figure \ref{fig:dust-profiles}, we plot dust extinction profiles extracted from \cite{2024A&A...685A..82E} maps toward the 28 Millennium targets with Zeeman measurements. The plotted profiles are the means of the 12 posterior samples of the 3D dust distribution. We observe that only a few Millennium Zeeman sight lines have dust peaks beyond $500 \mathrm{~pc}$ of the Sun. In orange, we highlight the distance range corresponding to the boundary of the Local Bubble wall, as modeled by \cite{2024ApJ...973..136O} based on \cite{2024A&A...685A..82E} maps. Many prominent dust peaks shown in Figure \ref{fig:dust-profiles} coincide with the boundary of the Local Bubble, indicating that some Zeeman measurements likely probe the nearby structure. We do not find a strong trend between the line-of-sight averaged $|B_{\text{LOS}}|$ and the distance to dust structures, although the few lowest $\langle|B_{\text{LOS}}|\rangle$ lines of sight appear to preferentially probe local ($d<200 \mathrm{~pc}$), high-Galactic-latitude gas with smaller total dust extinction, consistent with our findings in \S\ref{sec:environment}. 

\begin{figure}{}
    \centering
    \includegraphics[trim={0.cm 0cm 0cm 0cm}, clip, scale=0.75]{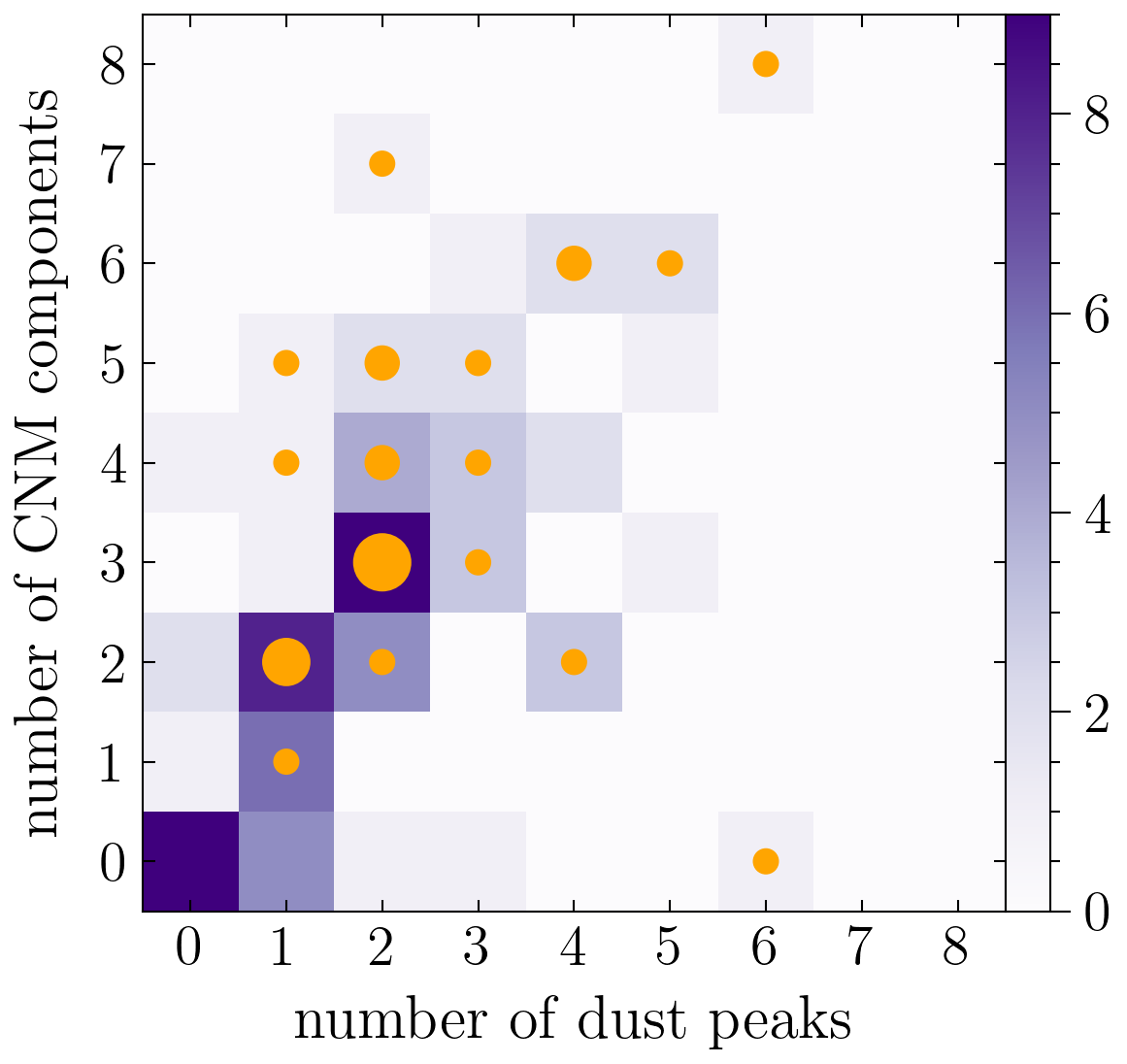}
    \caption{The number of dust peaks in \cite{2024A&A...685A..82E} extinction profiles against the number of Gaussian components in the HI optical depth spectrum.  The purple histogram includes all 80 Millennium sight lines, while the 28 Millennium sight lines with Zeeman measurements are plotted in orange circles scaled by the sight line count, which ranges from 1 to 6. }
    \label{fig:complexity}
\end{figure}

We assume that dust extinction is a good distance tracer of the bulk HI distribution and that CNM components are roughly associated with the peaks in $A_V$. This picture is supported by Figure \ref{fig:complexity}, where we find a strong correlation between the number of CNM components and dust peaks per sight line, both in the entire Millennium Survey (purple histogram) and toward the targets with Zeeman measurements (orange points). The dust peak count is derived using the peak finding algorithm described in \cite{2024ApJ...973..136O}, with the same smoothing parameter and minimum peak prominence of $10^{-4}$ in native units of the \cite{2024A&A...685A..82E} map. While some dust overdensities likely harbor multiple distinct cold HI components, Figure \ref{fig:complexity} indicates that the distance distribution of dust is generally informative of the distance distribution of HI absorption components. Based on this observation, we infer that at least some of the Zeeman components that are situated along the same line of sight probe structures separated by tens to hundreds of parsecs.

Given the general agreement in co-spectral Zeeman measurements for targets separated up to $7.5^{\circ}$ on the sky, we analyze their corresponding 3D dust profiles to infer the minimum physical distance between HI absorption components. In cases where both targets exhibit single-peaked dust profiles, we estimate the distance directly; for those with multiple peaks, we use the distance to the first peak to set a lower limit. The Zeeman components near $v = 4\,\textrm{km}\,\textrm{s}^{-1}$ toward the double radio galaxy sources 3C 225a and 3C 225b are in excellent agreement in magnetic field strength, spin temperature, and linewidth, but differ in optical depth. The dust distribution we extracted from 3D maps toward that region shows a simple profile with one prominent peak at $142$ pc, so we infer a physical separation between the two components of $0.25$ pc. The components near $-4\,\textrm{km}\,\textrm{s}^{-1}$ toward 3C 310 and 3C 315 also share similarities in spectral properties. The dust profiles of 3C 310 and 3C 315 have a single dominant peak at $\approx 105$ pc, corresponding to a $3.7\,$pc separation between the Zeeman components. Toward 3C 132 and 3C 133, the dust profiles are more complex, but their first prominent peaks both occur at $155$ pc. Assuming this is the peak associated with the absorbers, the minimum separation between the HI components is $7.6$ pc. While 3C 75 and 3C 78 lack co-spectral HI components, their dust distributions span a broad range of 100--400 pc with peaks at similar distances. The change in velocity and magnetic field direction could be a signature of interesting dynamics in the region, which lies in proximity to the wall of the Orion-Eridanus superbubble. The remaining target pairs either show complex profiles or do not contain high signal-to-noise Zeeman measurements, but assuming a minimum fiducial distance to structures of $100$ pc and an angular separation of $7^{\circ}$ implies a minimum physical separation of $12$ pc. In summary, we have examined the coherence between Zeeman-measured $B_{\text{LOS}}$ in velocity and position space and found tentative evidence that the magnetic field probed by Zeeman measurements in cold HI can be coherent on scales of at least $7.6$ pc.

\section{Discussion}\label{sec:discussion}

\subsection{Do HI Filaments Trace the Inclination of the Zeeman-Probed Magnetic Field?}

In this paper, we investigated the correlation between HI filament dispersion in narrow velocity channel maps of HI emission and Zeeman measurements in HI absorption.  We hypothesized that higher filament dispersion might be indicative of a magnetic field that is more parallel to the observer's line of sight, and therefore be associated with stronger Zeeman measurements of the line-of-sight field strength. Even in the case where the HI filament dispersion perfectly traces the magnetic field inclination angle, our expectation for the tightness of this correlation depends critically on the unknown distribution of the total field strength across clouds. If the CNM permits a narrow range of $B_{\text{TOT}}$, e.g., in the extreme case, PDF($B_{\text{TOT}}) \sim \delta(B_0-B_{\text{TOT}})$, we expect the correlation to tighten. If the distribution of $B_{\text{TOT}}$ is broad, there is a strong degeneracy between the field strength and orientation. In this scenario, a fully line-of-sight oriented field will have a disordered HI morphology and will be observed to have $B_{\text{LOS}}$ with a broad spread representative of the intrinsic spread in $B_{\text{TOT}}$. 

Our main finding in Figure \ref{fig:Blos-vs-disorder} -- a weak but statistically significant correlation between Zeeman measurements and HI filament disorder -- is consistent with the broad PDF($B_{\text{TOT}}$) variant of the hypothesis that HI filament dispersion is sensitive to magnetic field inclination. In this scenario, low values of $\text{Var}(\theta_{\text{HI}})$ correlate with low Zeeman measurements irrespective of $B_{\text{TOT}}$  because the field is dominantly oriented in the plane-of-sky, while high values of $\text{Var}(\theta_{\text{HI}})$ correlate with Zeeman measurements with large scatter due to the strength-orientation degeneracy. Additional spread might occur due to the high uncertainties in Zeeman measurements, velocity-blended HI morphologies, and the imperfect alignment of HI filaments with magnetic field lines, further weakening the correlation.

An important consideration in interpreting the $|B_{\text{LOS}}|$--$\mathrm{Var}(\theta_{\mathrm{HI}})$ correlation is how the physical scales probed by Zeeman measurements compare with those traced by HI filaments. We first detect the correlation when we compute $\text{Var}(\theta_{\text{HI}})$ on angular scales as small as 2.5$^{\circ}$, corresponding to 4.4 – 21.8 pc at distances of 100 – 500 pc where most Zeeman-traced gas is inferred to lie. Meanwhile, Zeeman measurements in absorption toward a point source are sensitive to a pencil-beam scale on the plane of the sky. Along the line of sight, however, these measurements probe length scales characteristic of the cold neutral medium, which may be typically on the order of a parsec \citep{1977ApJ...218..148M}. Zeeman absorption measurements also represent preferentially high optical depth components, which plausibly occupy the high end of the CNM size distribution, or may probe multiple CNM substructures distributed over a longer path length. \cite{2023MNRAS.521.5604H} studied synthetic Zeeman splitting of HI absorption lines in zoom-in simulations of molecular clouds and determined that these measurements typically probe scales of $\sim$10 pc. We conclude that Zeeman measurements may probe physical scales that are not so dissimilar from the scale on which we quantify filament dispersion. In the future, higher angular resolution HI data could provide sufficient HI filament statistics to compute $\mathrm{Var}(\theta_{\mathrm{HI}})$ on scales below 2.5$^{\circ}$, testing the observed trend at the most comparable scales. 

On the other hand, even if the scales directly probed by absorption components and HI filaments differ, a degree of coherence in the magnetic field inclination could persist over the range of scales represented by the two measurements. In our analysis of the spatial and velocity correlations between Zeeman measurements (\S\ref{sec:structure}), we observe that CNM components at similar velocities toward radio sources with angular separations as large as $7.5^{\circ}$ agree in the strength and direction of $B_{\text{LOS}}$. In sight lines where the mapping to dust structures is straightforward, we infer that such pairs of components are separated by at least a few parsecs. This coherence in independent Zeeman measurements on the scale of a few parsecs is also an agreement with observations of starlight polarization in parsec-scale, diffuse, intermediate-latitude fields, which exhibit highly ordered polarization angle structure \citep{2024AJ....168...47A}. HI emission filaments, which trace the magnetic field orientation on the plane of the sky, also appear to exhibit degree-scale coherence. Putman et al. (2025, in prep) found that the plane-of-sky orientation of HI emission filaments in the GALFA-HI survey remains correlated over a few degrees, in agreement with the result that the $|B_{\text{LOS}}|$--$\mathrm{Var}(\theta_{\mathrm{HI}})$ correlation investigated in this work persists when $\mathrm{Var}(\theta_{\mathrm{HI}})$ is computed on angular scales of 8$^{\circ}$. 

If $\mathrm{Var}(\theta_{\mathrm{HI}})$ is indeed sensitive to the inclination angle of the magnetic field, then systematic variations in inclination angle across the Arecibo footprint can explain the observed $|B_{\text{LOS}}|$--$\mathrm{Var}(\theta_{\mathrm{HI}})$ correlation, as shown in Figure \ref{fig:hist-AV-split}. For instance, we observe that high Galactic latitude sight lines exhibit a more filamentary morphology and lower $|B_{\text{LOS}}|$ than regions at intermediate latitudes. Many of the high-latitude measurements trace nearby ($<200$ pc) cold gas of substantial column density at distances consistent with the Local Bubble wall. It has been suggested that the Local Bubble hosts a magnetic field predominantly tangential to its surface, due to the compression of field lines by the supernova-driven expansion that formed the cavity \citep{2018A&A...611L...5A, 2025ApJ...988..191O}, which would result in lower Zeeman measurements towards high-latitude sight lines dominated by the Local Bubble gas. Over the high-latitude sky, and on the comparable angular scale of 160$^{\prime}$, \cite{2019ApJ...887..159H} observed an imprint of the magnetic field inclination angle on the unpolarized dust emission, which is a line-of-sight integrated measure, and thus sensitive to the inclination of the mean line-of-sight magnetic field. The measurements of $|B_{\text{LOS}}|$ could carry an imprint of this mean inclination while still exhibiting cloud-scale variations in the magnetic field direction. Because of the velocity localization, HI might be a more sensitive tracer of the local magnetic field geometry than dust emission, which, as a line-of-sight integrated quantity, is sensitive to magnetic field structure both within and beyond the Local Bubble wall \citep{2024ApJ...973...54H}. 

In summary, $|B_{\text{LOS}}|$--$\mathrm{Var}(\theta_{\mathrm{HI}})$ could arise due to the sensitivity of $\mathrm{Var}(\theta_{\mathrm{HI}})$ to the magnetic field inclination angle, either because Zeeman measurements and HI filaments trace similar physical scales, or because both probe a magnetic field coherent over a broader range of scales. Our analysis suggests HI filaments might be sensitive to systematic differences in the mean inclination angle along the line of sight, for instance, by returning lower $|B_{\text{LOS}}|$ and $\mathrm{Var}(\theta_{\mathrm{HI}})$ in sight lines dominated by the Local Bubble. Whether filament statistics can ultimately be combined with Zeeman data to recover $B_{\text{TOT}}$ at each velocity and position remains an open question that will require a larger, higher‑S/N sample of Zeeman measurements, and possibly still‑higher‑resolution HI maps to obtain robust filament statistics on sub‑degree scales.

\subsection{Alternative Interpretation: Impact of the Galactic Environment}

\begin{figure}{}
    \centering
    \includegraphics[trim={0.1cm 0cm 0cm -0.3cm}, clip, scale=0.62]{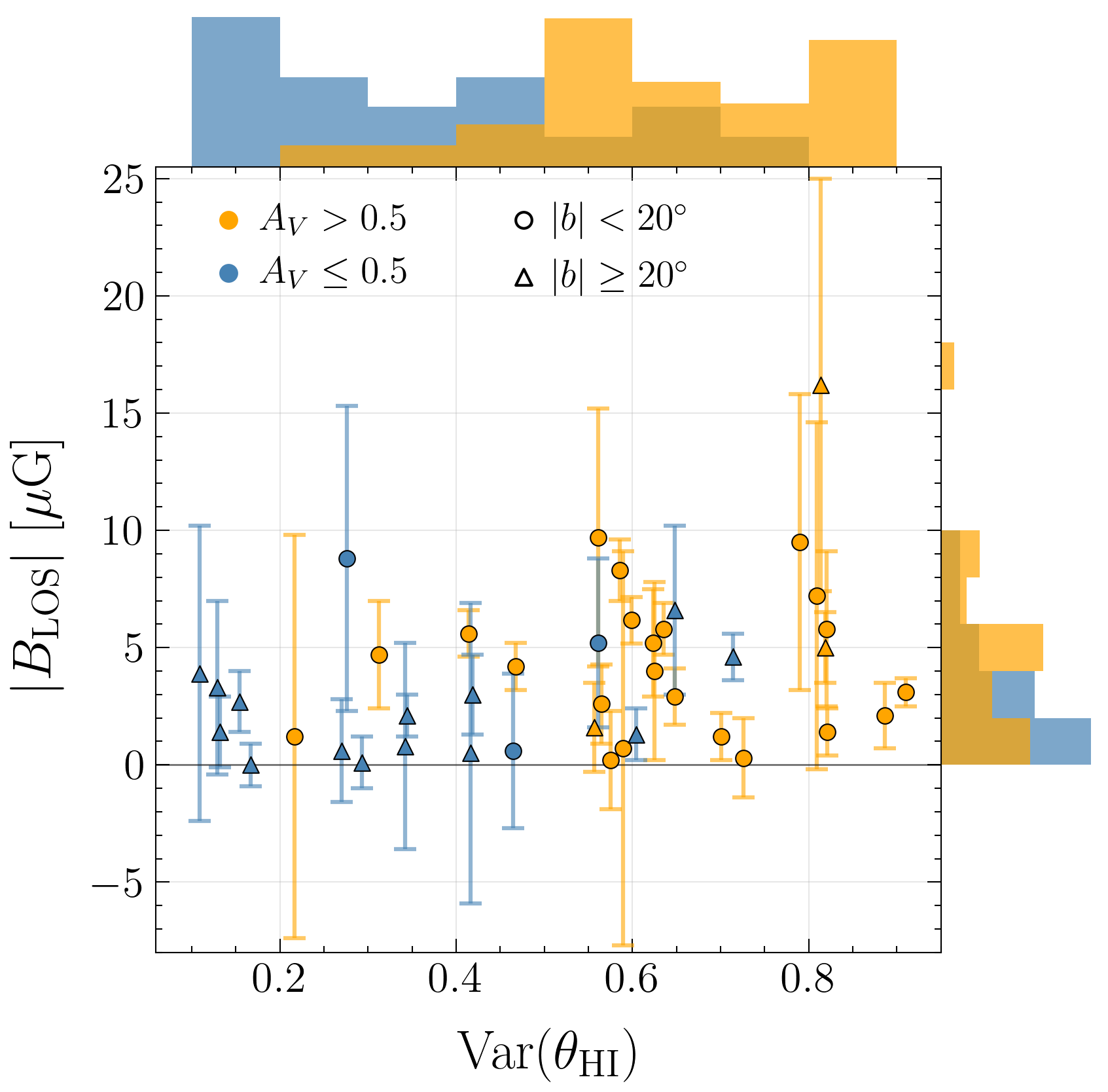}
    \caption{As in Figure \ref{fig:Blos-vs-disorder}, we plot Zeeman measurements $|B_{\text{LOS}}|$ against the circular variance of HI filament orientations $\text{Var}(\theta_{\mathrm{RHT}})$ computed in a $d_{\text{POS}} = 2.5^{\circ}$ region around each radio source. We use colors and symbols to split the sample by $A_V$ and Galactic latitude $b$, and observe that either split results in subsamples that largely occupy different regions of the plot. This suggests that systematic variations in magnetic field inclination or total field strength with $A_v$ or $b$ could drive the correlation between $|B_{\text{LOS}}|$ and $\text{Var}(\theta_{\mathrm{RHT}})$. }
    \label{fig:hist-AV-split}
\end{figure}

Figure \ref{fig:hist-AV-split} shows that the Galactic latitude dependence attributable to the large-scale inclination of the magnetic field coincides with a similar dependence on $A_V$. In Figure \ref{fig:sightline-blos-AV}, we observed that regions with a higher dust extinction, particularly in the vicinity of the Gould Belt, harbor some of the higher Zeeman measurements in the sample. This leads to an alternative interpretation where the $|B_{\text{LOS}}|$--$\mathrm{Var}(\theta_{\mathrm{HI}})$ correlation arises due to intrinsic differences in the total, rather than line-of-sight projected, magnetic field strength across Galactic environments. The corresponding change in HI morphology might reflect variations in magnetic field properties, such as differences in the Alfvén Mach number or a reduced alignment between HI filaments and the magnetic field, but it could also result from factors unrelated to the magnetic field that correlate with $A_V$.

This interpretation suggests that the PDF($B_{\text{TOT}}$) varies across the Galactic environment. In \S\ref{sec:environment}, we used different quantities to characterize the environment, including HI column density, dust extinction, and molecular gas tracers, and found that lines of sight with $A_V<0.5\,\mathrm{mag}$, or $N_{\mathrm{HI}}<7 \times 10^{20} \textrm{~cm}^{-2}$, have lower line-of-sight averaged $|B_{\text{LOS}}|$ than the rest of the sample. While this observation could hint at a potential relationship between $B_{\text{TOT}}$ and density, a similar trend has not been previously observed on the level of individual absorption components. \cite{2010ApJ...725..466C} used spin temperature $T_s$ to derive the volume density of hydrogen atoms $n_{\text{H}}$ assuming a constant interstellar pressure and found no evidence for the growth of the magnetic field with $n_{\text{H}}$ in the HI regime. This could be consistent with Figure \ref{fig:sightline-blos-AV} if the characteristics of the environment, rather than the volume density of individual HI components, drive differences in the total magnetic field strength. For example, 4 high-Galactic-latitude Millennium targets (3C 315, 3C 318, 3C 333, 3C 310) are in close proximity to a nearby ($\approx$ 100 pc) molecular cloud Eos \citep{2025NatAs...9.1064B}. The cloud was discovered recently via $\mathrm{H}_2$ fluorescence, is almost entirely CO-dark, and has never formed stars \citep{2025MNRAS.540L.109S}. On average, these 4 Millennium sources have lower Zeeman magnetic field strengths than sources intersecting the Gould Belt, which hosts star-forming regions where the magnetic field might be shaped by enhanced turbulence or stellar feedback \citep[e.g.,][]{2018A&A...609L...3S}.


Beyond the dependence on $A_V$ and $N(\text{HI})$, we have not identified a physical parameter that can uniquely explain the environmental difference. For instance, we did not find a strong dependence of the sight line-averaged and individual Zeeman field strengths on the presence of $\mathrm{H}_2$ traced using OH and CO spectral lines, although these tracers might not always reveal the presence and extent of $\mathrm{H}_2$, as in the case of Eos. Further, there is no statistically significant correlation between $|B_{\mathrm{LOS}}|$ or $\mathrm{Var}(\theta_{\mathrm{HI}})$ and the turbulent velocity dispersion of HI lines $\sigma_{v, \textrm{NT}}$. Future studies focusing on individual regions and a larger statistical sample of Zeeman measurements may help to disentangle the drivers of the observed variations in $|B_{\mathrm{LOS}}|$.

\section{Conclusions}\label{sec:conclusions}

In this work, we examined the connection between the dispersion of HI emission filaments and Zeeman measurements in HI absorption and found a weak, positive correlation between the two measures. The correlation is statistically significant at a $p \approx 0.01$ level. We considered two interpretations of the observed trend: in the first scenario, Zeeman measurements are sensitive to the inclination of the magnetic field probed by the HI filaments, either because both tracers probe similar physical scales or because the magnetic field orientation is coherent over the range of scales they sample. Our analysis shows that both $|B_{\text{LOS}}|$ and $\mathrm{Var}(\theta_{\mathrm{HI}})$ vary systematically across the sky, with lower values observed in regions dominated by Local Bubble gas, consistent with a magnetic field oriented preferentially in the plane of the sky. We also considered a second, alternative scenario, in which the total magnetic field strength changes across the Galactic environment in conjunction with HI filament dispersion, resulting in higher $|B_{\text{LOS}}|$ in sight lines with higher dust extinction. These two interpretations are not mutually exclusive and might be complementary, with inclination effects influencing high-latitude measurements and environmental factors influencing molecular cloud envelopes. To firmly establish whether HI filament morphology in combination with Zeeman measurements can be used to reconstruct the total magnetic field strength, $B_{\text{TOT}}$, a larger sample of sensitive HI Zeeman measurements probing diverse sight lines is needed.

Below we summarize our results:

\begin{itemize}
    \item We present new FAST data of Zeeman splitting in HI absorption toward three continuum radio sources from the Millennium Survey. We find magnetic field strength consistent with Arecibo observations for all nine absorption components, reproducing a $5\sigma$ Arecibo detection toward 3C 409 at a $4\sigma$ level. 
    \item We use the Rolling Hough Transform to quantify the disorder of HI filaments detected in narrow-channel GALFA-HI maps at the positions and velocities of Zeeman measurements in HI absorption. In a subsample of 42 spectrally distinct components, we find a statistically significant, weak correlation between filament dispersion and $|B_{\text{LOS}}|$, which is independent of algorithmic choices, persists to large angular scales, and is only weakly sensitive to the velocity of the narrow-channel HI map.  
    \item We investigate whether the correlation can be driven by an environmental difference in the total magnetic field strength. We find that sight lines with higher $A_V$ and $N(\text{HI})$ host higher mean $\langle|B_{\text{LOS}}|\rangle$, which can be attributed either to larger $B_{\text{TOT}}$ or smaller inclination angle. We also investigate the connection between $|B_{\text{LOS}}|$ and the presence of molecular gas traced by CO emission and OH absorption, but do not find a consistent association. 
    \item We find a strong correlation between the number of HI absorption components and the number of peaks in 3D dust maps toward Millennium targets, demonstrating that 3D dust maps are informative of cold HI distance distribution. Using 3D dust maps, we infer that most Zeeman measurements probe nearby structures ($100-500$ pc), some at distances consistent with the Local Bubble wall. 
    \item We examine the coherence of Zeeman measurements as a function of velocity and plane-of-sky separation between radio sources. We find enhanced coherence in the sample of co-spectral (within $3\,\textrm{km}\,\textrm{s}^{-1}$) components with small (up to $7.9^{\circ}$) angular separations. For pairs of components that agree in the strength and direction of the magnetic field, we estimate a physical separation of a few parsecs based on the 3D dust maps, suggesting HI absorption Zeeman measurements trace magnetic fields that maintain coherence across parsec-scale regions.
    \item We propose that the observed correlation between Zeeman measurements and HI filament disorder arises either due to systematic differences in magnetic field inclination (e.g. due to the Local Bubble geometry) or because of enhancement in the magnetic field strength in regions of high dust extinction. More high-S/N Zeeman measurements across diverse Galactic environments are needed to differentiate between these interpretations. 
\end{itemize}

\begin{acknowledgments}
M.N. thanks Adam Mantz, Philipp Frank, Sergio Martin-Alvarez, Claire E. Murray, Minjie Lei, Mehrnoosh Tahani, Juan D. Soler, Antoine Marchal, Amit Seta, and Iryna Butsky for helpful conversations. This work was supported by the National Science Foundation under grants No.\ AST-2106607 and AST-2441452, and by NASA award 80NSSC23K0972. S.E.C.\ additionally acknowledges support from an Alfred P. Sloan Research Fellowship. Tao-Chung Ching is a Jansky Fellow of the National Radio Astronomy Observatory. 

The authors acknowledge Interstellar Institute's program "ii7" and the Paris-Saclay University's Institut Pascal for hosting discussions that nourished the development of the ideas behind this work. This work also benefited from the conference ``Structure and polarization in the interstellar medium: A Conference in Honor of Prof. John Dickey", a hybrid meeting hosted jointly at Stanford University and at the Australia Telescope National Facility in February 2025. We acknowledge support from the National Science Foundation (NSF Award No. 2502957), from the Kavli Institute for Particle Astrophysics and Cosmology, from the Commonwealth Scientific and Industrial Research Organisation, and from the Australian Research Council.

This publication utilizes data from the Galactic ALFA HI (GALFA HI) survey dataset obtained with the Arecibo L-band Feed Array (ALFA) on the Arecibo 305 m telescope. The Arecibo Observatory is operated by SRI International under a cooperative agreement with the National Science Foundation (AST-1100968), and in alliance with Ana G. Méndez-Universidad Metropolitana, and the Universities Space Research Association. The GALFA HI surveys have been funded by the NSF through grants to Columbia University, the University of Wisconsin, and the University of California.

\end{acknowledgments}

\software{Astropy \citep{2013A&A...558A..33A}, ChatGPT \citep{openai2022chatgpt}, CMasher \citep{2020JOSS....5.2004V}, dustmaps \citep{2018JOSS....3..695G}, healpy \citep{2019JOSS....4.1298Z}, matplotlib \citep{Hunter:2007}, numpy \citep{harris2020array}, pandas \citep{reback2020pandas}, RHSTK, Rolling Hough Transform \citep{2014ApJ...789...82C}, scienceplots \citep{SciencePlots}, scipy \citep{2020NatMe..17..261V}, seaborn \citep{Waskom2021}}

\bibliography{references}
\bibliographystyle{aasjournalv7}

\end{document}